\documentclass[
aps,
prb,
reprint,
superscriptaddress,
amsmath,
amssymb,
longbibliography,
]{revtex4-1}

\usepackage{graphicx}% Include figure files
\usepackage{dcolumn}% Align table columns on decimal point
\usepackage{bm}% bold math
\usepackage{mhchem}% Chemial formula
\usepackage{natbib,hyperref}% add hypertext capabilities
\hypersetup{colorlinks=true,linkcolor=blue,citecolor=blue,urlcolor=blue}
\usepackage{color}
\usepackage{ulem}
\usepackage{float}

\renewcommand{\vec}[1]{\mathbf{#1}}
\renewcommand{\emph}[1]{\textit{#1}}

\begin{document}

\title{Near Unity Optical Spin Polarization in GaSe Nanoslabs}

\author{Yanhao Tang}
\author{Wei Xie}
\affiliation{Department of Physics and Astronomy, Michigan State University, East Lansing, MI 48824, USA}
\author{Krishna C. Mandal}
\affiliation{Department of Electrical Engineering, University of South Carolina, Columbia, SC 29208, USA}
\author{John A. McGuire}
\author{C. W. Lai}
\email{cwlai@msu.edu}
\affiliation{Department of Physics and Astronomy, Michigan State University, East Lansing, MI 48824, USA}

%\date{\today}

\begin{abstract}
{We report nearly complete preservation of ``spin memory" between optical absorption and photoluminescence (PL) in nanometer slabs of GaSe pumped with 0.2 eV excess energy. At cryogenic temperatures, the initial degree of circular polarization ($\rho_0$) of PL approaches unity, 
with the major fraction of the spin polarization decaying with a time constant $>$500 ps in sub-100-nm GaSe nanoslabs. Even at room temperature, $\rho_0$ as large as 0.7 is observed, while pumping 1 eV above the band edge yields $\rho_0$ = 0.15. Angular momentum preservation for both electrons and holes is due to the separation of the non-degenerate conduction and valence bands from other bands. In contrast to valley polarization in atomically thin transition metal dichalcogenides, here optical spin polarization is preserved in nanoslabs of 100 layers or more of GaSe.} 
\end{abstract}

%\keywords{}

\maketitle

Solid-state systems exhibiting high spin polarization and long spin relaxation time are desirable for spintronic applications. Various semiconductors have been studied for creation of long-lived non-equilibrium spin populations and coherences \cite{dyakonov1984,zutic2004,dyakonov2008,awschalom2013}. The most extensively studied system is gallium arsenide (GaAs). However, optically pumped electron spin polarization is limited to 1/2, while the maximal degree of circular polarization of photoluminescence is 1/4 \cite{dyakonov1984,dyakonov2008}, owing to the degenerate heavy- and light-hole valence bands and sub-ps hole spin relaxation. Doping \cite{kikkawa1997} or quantum confinement \cite{ohno1999} has been used to quench \emph{electron} spin relaxation. Unity electron spin polarization can be achieved in heterostructures where heavy- and light-hole energy degeneracy is lifted by quantum confinement or strain. Still, near-resonant optical excitation is necessary to avoid transitions involving both heavy- and light-hole bands. In analogy to spin polarization, valley polarization has been demonstrated in monolayer transition metal dichalcogenides (TMDs) with potential applications exploiting both spin and valley degrees of freedom.In monolayer TMDs with broken inversion symmetry, a direct gap emerges at the corners ($K$ points) of the Brillouin zone, enabling valley-dependent inter-band transitions under circularly polarized optical excitation \cite{yao2008,cao2012,zeng2012}. Furthermore, the substantial spin-splitting of valence bands at the band edges due to spin-orbit interaction originating from the $d$ orbitals of TM ions has led to recent reports of long hole spin and valley lifetimes \cite{xiao2012,mak2012}, valley exciton polarization and coherence \cite{xu2014,zhu2014}, circularly polarized electroluminescence \cite{zhang2014}, and valley Hall effect \cite{mak2014}. Indeed, circularly polarized PL was observed in single- and bi-layer TMDs \cite{cao2012,zeng2012,mak2012,xu2014,zhu2014} with steady-state near-resonant circularly polarized excitation. However, time- and polarization-resolved PL measurements suggest that, at least in \ce{MoS2}, circularly polarized PL can result from sub-10-ps recombination and valley (spin) lifetimes rather than an intrinsically long-lived valley or hole spin polarization \cite{cao2012,lagarde2014,glazov2014}. Additionally, emission at the direct gap becomes dominant only at the monolayer level \cite{splendiani2010,mak2010,tonndorf2013}.

\begin{figure}[]
\includegraphics[width=0.5 \textwidth]{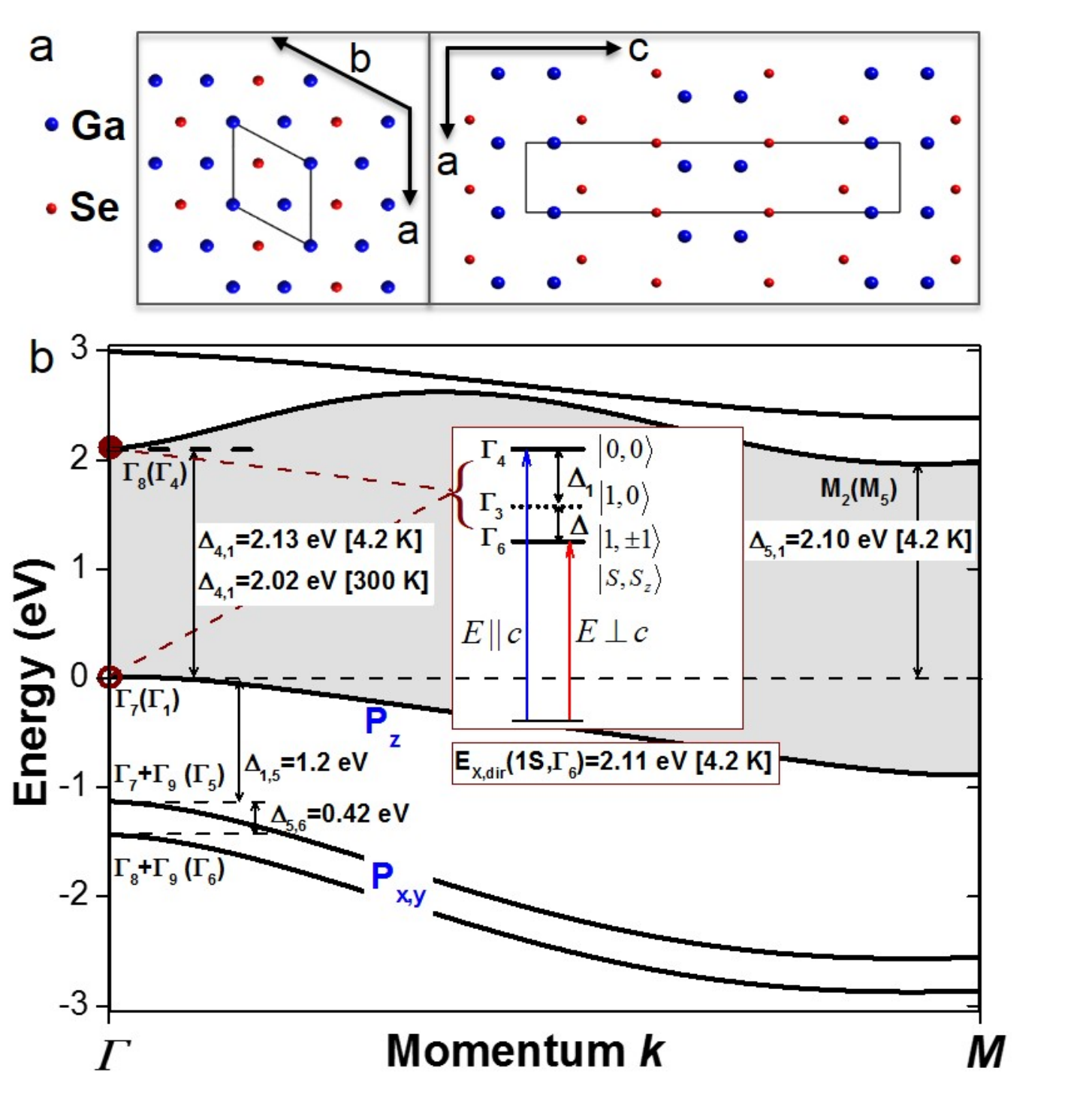}
\caption{\label{fig:band}\textbf{Band structure and selection rules.} 
(\textbf{a}) Crystal structure of $\epsilon$-GaSe showing $ABA$ stacking of the individual layers, space group $D_{3h}^{1}/P\bar{6}m2$ (\#187) and point group $D_{3h}$. An individual layer consists of four planes of \ce{Se-Ga-Ga-Se}, with the \ce{Ga-Ga} bond normal to the layer plane arranged on a hexagonal lattice and the Se anions located in the eclipsed conformation when viewed along the $c$-axis. The inner solid boxes represent a unit cell. (\textbf{b}) Sketches of the band structure of $\epsilon$-GaSe at the $\Gamma$ point and the representations to which the states at the $\Gamma$-point belong with (without) spin-orbit interaction. (inset) Direct-gap excitons and selection rules.}
\end{figure}

Here, we demonstrate GaSe as a promising material for generating and preserving a high degree of spin polarization. The unique bandstructure (Fig. \ref{fig:band}) of the group-III monochalcogenides removes degeneracies between orbital states and thereby allows generation and preservation of a high degree of spin polarization \cite{gamarts1977,ivchenko1977}. We investigate the photoluminescence of excitons in sub-100-nm to 1000-nm thick GaSe slabs (nanoslabs) under non-resonant circularly polarized optical excitation. Polarized time-dependent PL in GaSe reveals a high initial degree of circular polarization $\rho_0>0.9$ when excited with excess energy from about 0.1 to 0.2 eV. High $\rho_0$ is indicative of spin preservation of optically active carriers during the optical absorption-cooling process. Unlike TMDs, GaSe has a quasi-direct gap in the bulk form \cite{mooser1973,capozzi1993}, making it a potentially versatile material in which strong emission occurs and light-matter coupling can be controlled from bulk to nanoscale thicknesses.

%\subsection*{Results}
%\subsection{Band structure, optical orientation and selection rules}
The physical properties of \ce{GaSe} are largely determined by those of a single layer (Fig. \ref{fig:band}) due to the high electron density within the layer and weak interlayer interaction \cite{mooser1973,schluter1976}. The Se-Ga bond tilts 29$^\circ$ out of the layer plane so that Se 4$p_{x,y}$ electrons experience much greater Coulomb attraction with \ce{Ga} cations than Se 4$p_z$ electrons do. This results in a large negative crystal field ($\Delta_{cr} \approx$ -1.4 eV) compared to wurtzite-type materials ($\Delta_{cr} \sim$ +0.05 eV) so that the Se 4$p_{x,y}$ states lie 1.2 and 1.6 eV below the Se 4$p_z$ state (Fig. \ref{fig:band}b). 
Consequently, near the $\Gamma$ point ($\vec{k}=0$) only light with $\vec{E} \parallel c$ can excite dipole transitions between the $p_z$-like uppermost valence band and $s$-like lowest conduction band. Experimentally, the absorption coefficient for $\vec{E} \perp c$ is about $10^3$ cm$^{-1}$, a factor of 30 smaller than for $\vec{E} \parallel c$ \cite{le-toullec1977}. The transition for $\vec{E} \perp c$ is weakly dipole-allowed owing to band mixing induced by strong spin-orbit interaction ($\Delta_{so} \, \approx$ 0.44 eV) and weak interlayer coupling. The optical pumping and selection rules near $k=0$ are best understood in an exciton (two-particle) picture [binding energy $\sim$ 20-30 meV] \cite{mooser1973} as detailed in Supplementary Information. The essential feature is that the lowest states correspond to total exciton spin $S=1$ with $z$-component $S_z=\pm1$ and can be excited by light with wave vector $\vec{k} \parallel c$ ($\vec{E} \perp c$). We have exploited these selection rules to investigate the spin dynamics of GaSe under non-resonant circularly polarized optical excitation as a function of slab thickness $d_L$.

%\section{Experimental procedures}
Time- and polarization-resolved PL measurements (see Methods) allow us to separately determine the recombination time, the initial spin orientation, and the spin relaxation time. Polarization-resolved PL measurements are performed under excitation with energy about 0.1 to 0.2 eV above the GaSe band gap. The band-edge exciton PL emission at room temperature is near 620 nm (2.0 eV), independent of thickness, under 594 nm (2.087 eV) excitation. Additionally, we observe that the quantum yield of luminescence is greatly suppressed in sub-50-nm thick samples (Supplementary Fig. S2). In this paper, we focus on GaSe nanoslabs ranging from about 90 to 2000 nm thick. 

\begin{figure}[]
\includegraphics[width=0.5 \textwidth]{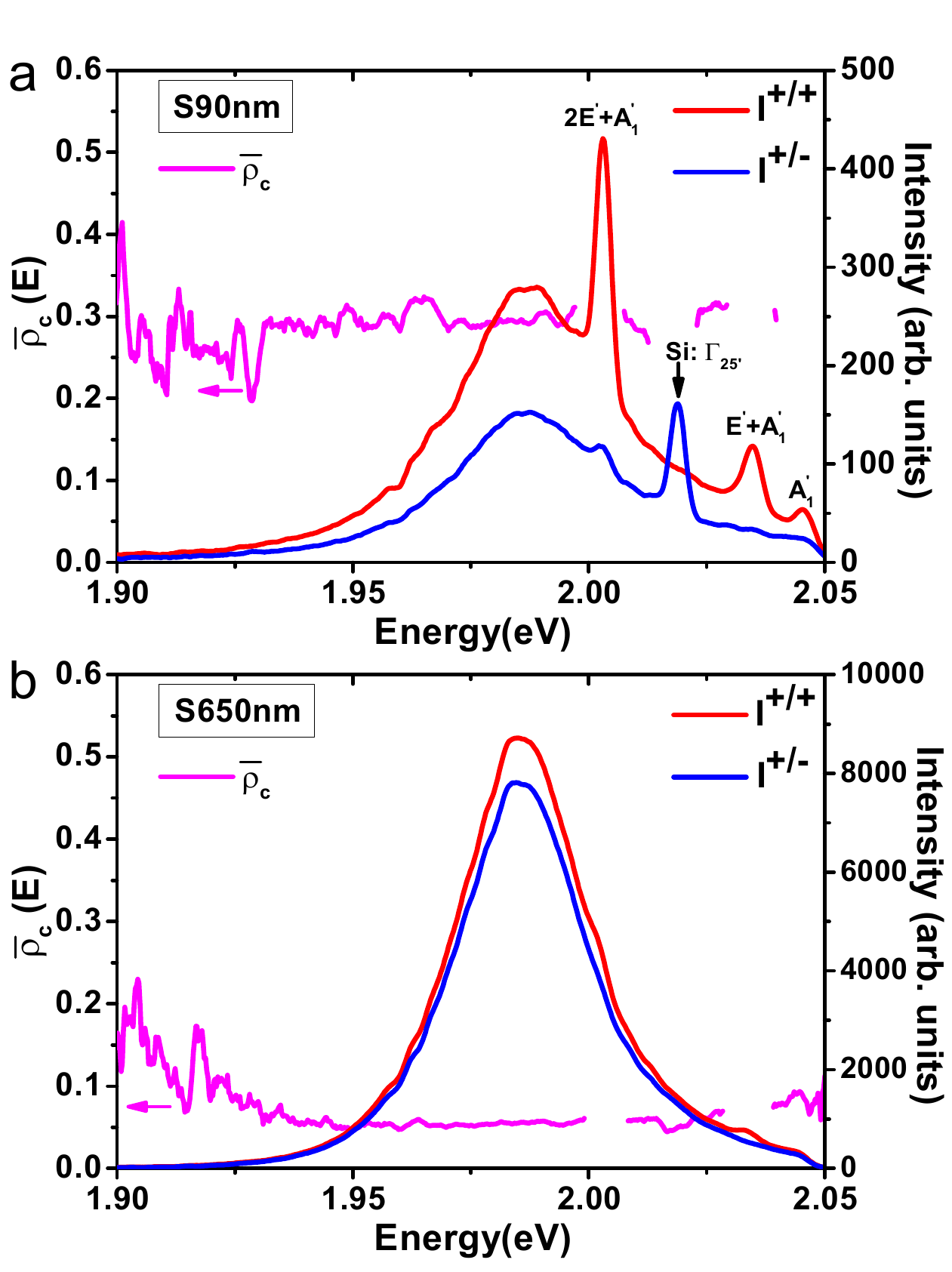}
\caption{\label{fig:PL300K} \textbf{Polarized time-integrated PL spectra at T = 300 K.} Time-integrated PL spectra [$I^{+/+}(E)$(co-circular excitation and detection, red) and $I^{+/-}(E)$(cross-circular excitation and detection, blue)] and degree of circular polarization $\bar{\rho}_c(E)=\frac{I^{+/+}(E)-I^{+/-}(E)}{I^{+/+}(E)+I^{+/-}(E)}$ (magenta) of (\textbf{a}) 90-nm [S90nm] and (\textbf{b}) 650-nm thick [S650nm] GaSe samples under $\sigma^+$ excitation at pump flux $P$ = 0.5 $P_0$, where $P_0$ = $2.6\times10^{14}$  cm$^{-2}$ per pulse. At $P_0$, the photoexcited carrier density is about $3.4\times10^{17}$ cm$^{-3}$ ($2.7\times10^{10}$ cm$^{-2}$ per layer). Spectra under $\sigma^-$ excitation are symmetric with respect to those under $\sigma^+$ excitation (not shown). The labeled spectral peaks are one-phonon Raman mode $A'_{1}$ (310 cm$^{-1}$) and multi-phonon Raman modes $E'(LO)+A'_1$ [255 + 136 cm$^{-1}$] and $2E'(LO)+A'_1$ [510 + 136 cm$^{-1}$] in GaSe. The multi-phonon Raman modes ($\sim$394 and 648 cm$^{-1}$) become dominant in sub-100-nm thick GaSe and have not been reported in bulk GaSe. Raman shifts are calibrated against the 520-cm$^{-1}$ Raman line ($\Gamma_{25'}$ phonon mode) of the silicon substrate. 
}
\end{figure}

\begin{figure*}[]
\includegraphics[width=1.0 \textwidth]{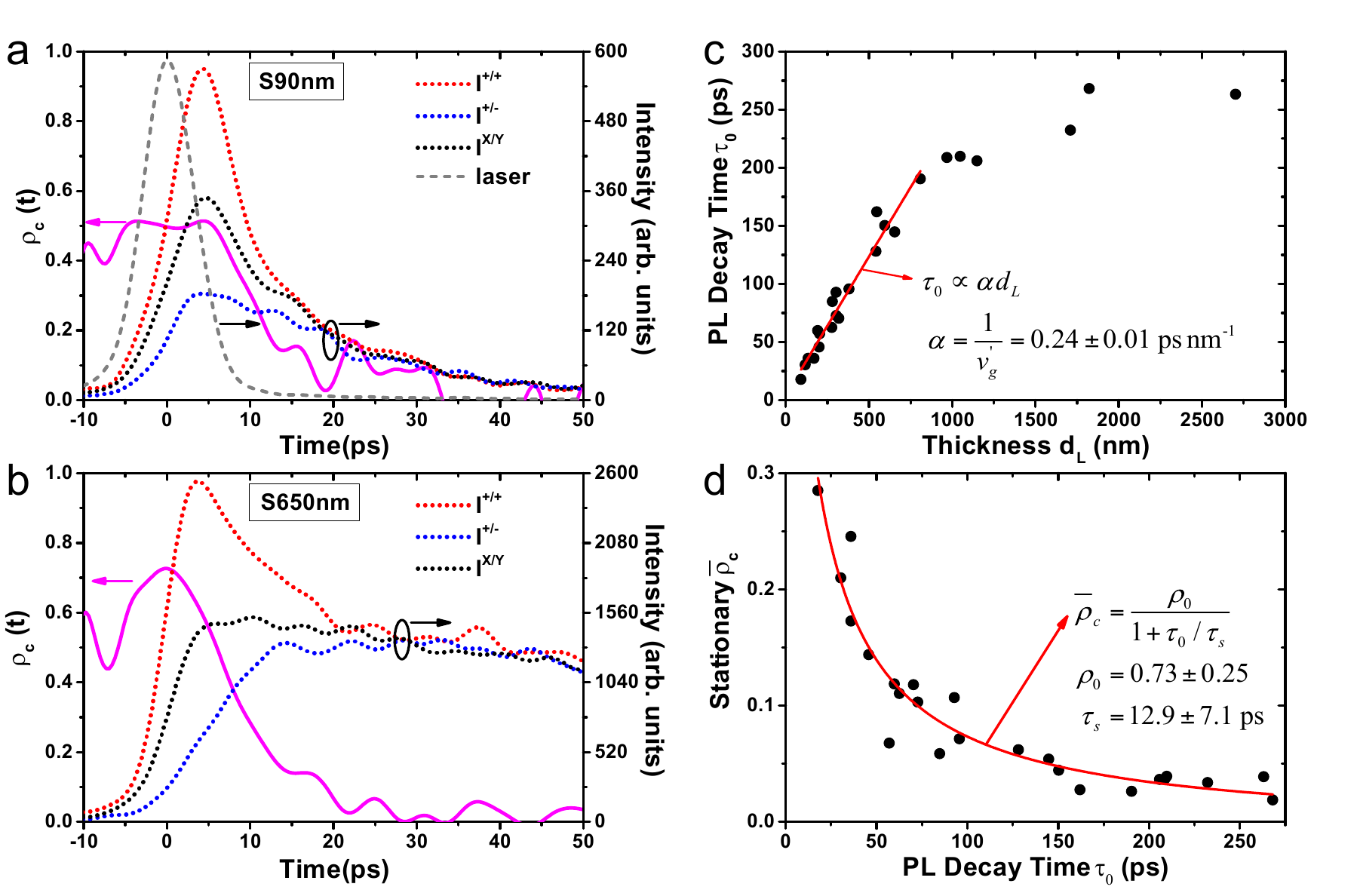}
\caption{\label{fig:TRPL300K}\textbf{Polarized time-dependent PL at T = 300 K.} (\textbf{a,b}) Time-dependent PL intensity [$I^{+/+}(t)$, $I^{+/-}(t)$, and $I^{X/Y}(t)$ (dashed red, blue, and black, respectively)] and degree of circular polarization $\rho_c(t)=\frac{I^{+/+}(t)-I^{+/-}(t)}{I^{+/+}(t)+I^{+/-}(t)}$ (solid magenta) of samples S90nm and S650nm under $\sigma^+$ or $\sigma^X$ excitation at $P$ = 0.5 $P_0$. The signal is spectrally integrated over the exciton PL peak. Time-zero is set to the peak of the instrument response function (grey dashed line) measured by the pump laser light scattered off the sample. (\textbf{c}) PL decay time constant ($\tau_0$) as a function of thickness ($d_L$) of the samples. $\tau_0$ decreases linearly with thickness for $d_L\lesssim$ 800 nm, yielding a slope of 0.24 ps/nm or an effective group velocity of 4.2 nm/ps. (\textbf{d}) The stationary (time-averaged) circular polarization near the exciton PL peak [$\bar{\rho}_c(E\approx1.98 \, \text{eV})$] versus $\tau_0$.
}
\end{figure*}

Upon absorption of circularly polarized light above the band-gap, the stationary (time-averaged) degree of circular polarization ($\bar{\rho}_c$) of luminescence represents photoexcited carrier spin memory. In Fig. \ref{fig:PL300K}, we show the polarized time-integrated PL spectra [$I(E)$] for two samples of thickness of 90 nm and 650 nm at room temperature. Across a range of sample thicknesses ($d_L$), time-integrated PL shows a pronounced increase of $\bar{\rho}_c$ from $\sim$0.05 in bulk to 0.3 in sub-100-nm nanoslabs. Corresponding polarized time-dependent PL are shown in Fig. \ref{fig:TRPL300K}a-b. The degree of circular polarization $\rho_c(t)$ decays within 10 ps, independent of the slab thickness and photoexcited carrier density. In contrast, the total (unpolarized) PL decays exponentially with a time constant ($\tau_0$) that increases linearly with thickness from about 20 ps to 250 ps for 90 nm $\lesssim d_L \lesssim$ 700 nm (Fig \ref{fig:TRPL300K}c). The fast PL rise and decay time constants in sub-100-nm GaSe nanoslabs are comparable to the $\sim$10-ps spin relaxation time, resulting in the preservation of a high spin polarization during the brief absorption-cooling-emission cycle \cite{nusse1997}. In Fig. \ref{fig:TRPL300K}d, we plot $\bar{\rho}_c$ as a function of PL decay time $\tau_0$ for \emph{all} samples studied. The stationary $\bar{\rho}_c$ is well described by
a simple model: $\bar{\rho}_c = {\rho_0}/{(1+\tau_0/\tau_s)}$, where $\tau_0$ is the lifetime of photoexcited carriers, and $\tau_s$ is the spin relaxation time. Fitting yields $\rho_0 = 0.73 \pm 0.25$ and $\tau_s=13 \pm 7$ ps, consistent with the time-resolved measurements. Therefore, the apparent enhancement of ``spin memory'' revealed in time-integrated PL is largely owing to a linear decrease of the PL decay time ($\tau_0$) with thickness.

\begin{figure}[]
\includegraphics[width=0.45 \textwidth]{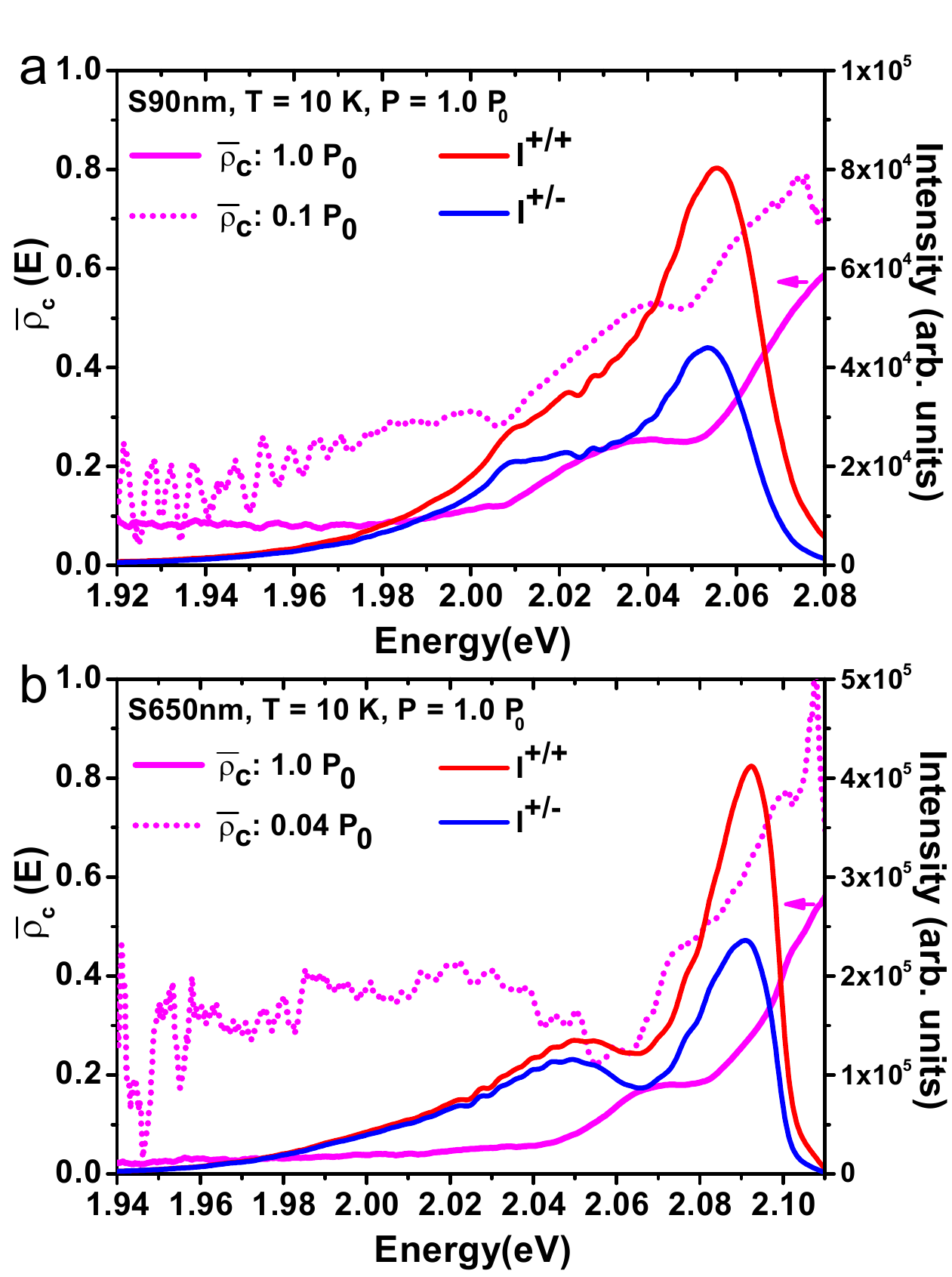}
\caption{\label{fig:PL10K}\textbf{Polarized time-integrated spectra at T = 10 K.} Time-integrated PL spectra [$I^{+/+}(E)$(co-circular, red) and $I^{+/-}(E)$(cross-circular, blue)] and degree of circular polarization $\bar{\rho}_c(E)$ of (\textbf{a}) S90nm and (\textbf{b}) S650nm under $\sigma^+$ excitation at pump flux $P$ = 1.0 $P_0$. $\bar{\rho}_c(E)$ at $P$ = 0.1 $P_0$ for S90nm and 0.04 $P_0$ for S650nm are shown as magenta dotted lines for comparison. $I(E)$ spectra at $P$ = 0.1 $P_0$ are shown in Fig. S3. Note that the exciton peak gradually red shifts from 590 nm (2.1 eV) to 620 nm (2.0 eV) in bulk ($> 1000$ nm) to 90-nm nanoslabs. This shift likely reflects enhanced exciton binding or Stokes shift, which is the subject of ongoing studies and beyond the scope of this paper.
}
\end{figure}

\begin{figure*}[]
\includegraphics[width=1.0 \textwidth]{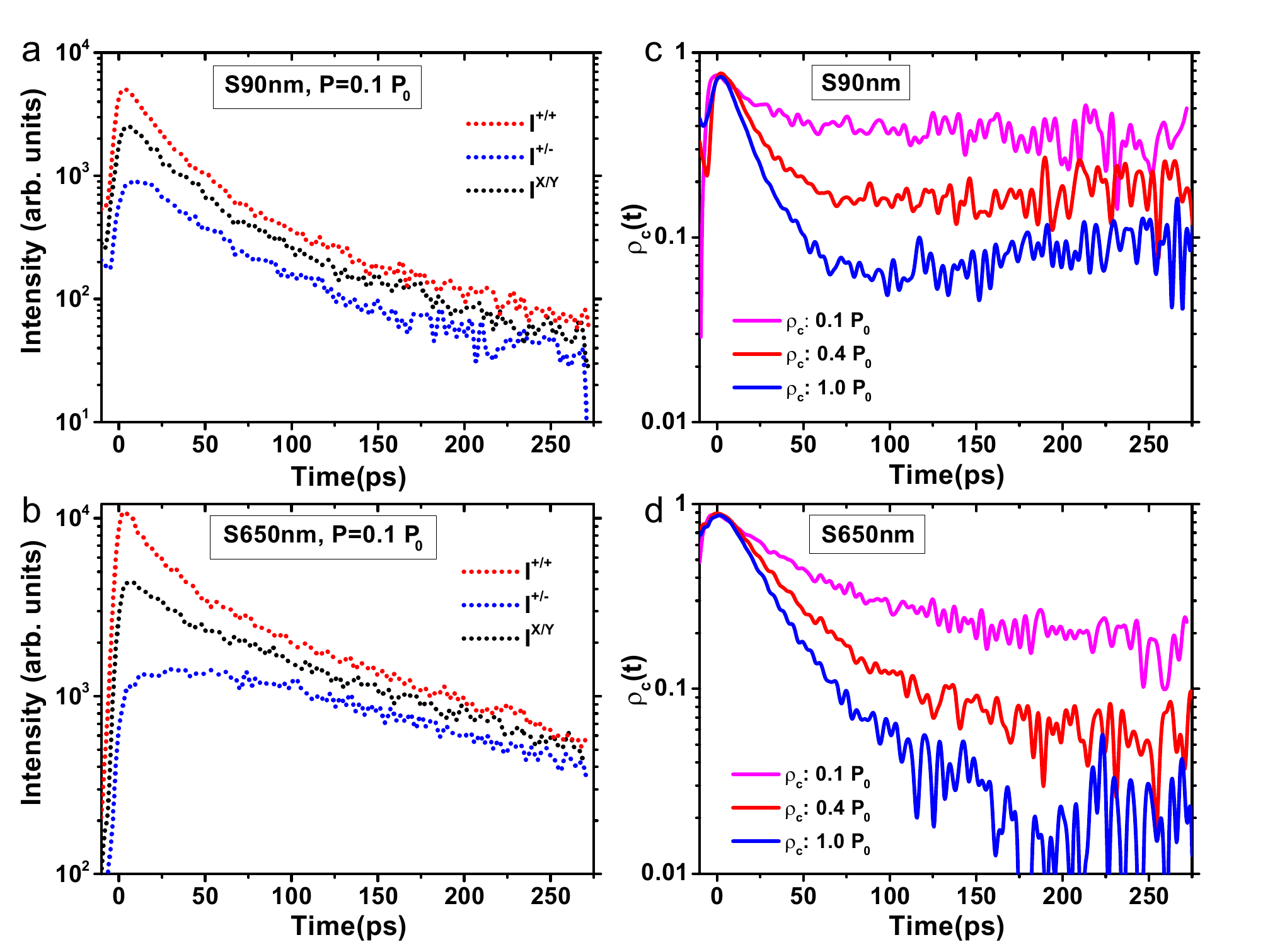}
\caption{\label{fig:TRPL10K}\textbf{Polarized time-dependent PL at T = 10 K.} (\textbf{a,b}) Time-dependent PL intensity [$I^{+/+}(t)$, $I^{+/-}(t)$, and $I^{X/Y}(t)$ (dashed red, blue, and black, respectively)] of S90nm and S650nm under $\sigma^+$ or $\sigma^X$ excitation at $P$ = 0.1 $P_0$. \textbf{(c,d)} $\rho_c(t)$ (magenta, red, and blue) of PL from S90nm and S650 under $\sigma^+$ excitation at a pump flux of $P$ = 0.1, 0.4, and 1.0 $P_0$.}
\end{figure*}

At cryogenic temperature (T = 10 K), the spin and PL decay dynamics with 560-nm (2.214) to 575-nm (2.156 eV) excitation slow markedly. For both 90-nm and 650-nm samples, the stationary $\bar{\rho}_c(E)$ surpasses 0.8 (approaching 1 at low photoexcited density) at the high-energy range of the exciton emission spectrum and decreases gradually to about 0.3 at low emission energies (Fig. \ref{fig:PL10K} and Supplementary Fig. S3 \& S4). Similar $\bar{\rho}_c(E)$ is observed for all samples (not shown). Meanwhile, time-dependent PL decay becomes bi-exponential with time constants $\tau_0' \, \approx$ 20-50 ps and $\tau_0'' \, \approx$ 150-200 ps (Fig. \ref{fig:TRPL10K}a-b) in all samples. In contrast, the PL rise time remains $<10$ ps, suggesting that the cooling and momentum relaxation of radiative photoexcited carriers are independent of temperature. Finally, time-dependent circular polarization $\rho_c(t)$ is also found to be bi-exponential, with decay time constants $\tau_s' \, \approx$ 30-40 ps and $\tau_s'' \, \gtrsim$ 500 ps (Fig. \ref{fig:TRPL10K}c-d). Assuming a high spin relaxation rate of non-thermal carriers during the cooling process, we reproduce quantitatively the bi-exponential decays of both PL intensity and polarization using a rate-equation model \cite{vinattieri1994} (see Supplementary Information). The shorter time constant $\tau_s'$ shows a weak power-law dependence on the photoexcited density ($n_X$) with $\tau_s' \propto n_X^{(-0.23\pm0.06)}$, while there is no identifiable dependence on thickness for either population decay or spin relaxation. 

To explore the limits of spin memory, we also performed experiments on the 650 nm sample under 3.0 eV excitation (Supplementary Fig. S5). We find that the initial $\rho_0$ at the band edge is 0.15, irrespective of temperature. The reduction of $\tau_s'$ with increasing carrier density and $\rho_0$ with increasing excitation energy suggest that the Elliot-Yafet (EY) mechanism plays a larger role than the Dyakonov-Perel (DP) mechanism \cite{dyakonov1984,pikus1984,wu2010} in the initial spin relaxation for non-thermal carriers at low temperature. 

%\subsection*{Discussion}
The generation of a substantial degree of spin polarization at the band edge after highly non-resonant excitation is a consequence of (1) a large degree of spin polarization of the initially created high-energy carriers and (2) spin relaxation rates that are slow compared to energy and momentum relaxation as well as the band-edge PL decay rates. All samples display a nearly instantaneous rise ($\sim$5 ps, resolution-limited) in the band-edge PL under excitation at excess energies $\lesssim$300 meV at room and cryogenic temperatures, indicating sub-10-ps momentum and energy relaxation of photoexcited carriers toward $k$ =0 ($\Gamma$ point). This time scale is consistent with the momentum scattering time of 5 ps deduced by Gamarts et al. \cite{gamarts1977,ivchenko1977}. Carriers thus spend limited time in high-momentum states where spin relaxation is fastest. In contrast, at 3.0 eV excitation, carriers take markedly longer to dissipate 1 eV excess energy, especially at T = 10 K. This increase of energy relaxation time partially accounts for the reduced initial degree of spin polarization observed under 3.0 eV excitation. 

The decrease of PL decay $\tau_0$ with thickness for $d_L<$ 700 nm at $T=300$ K means that a greater fraction of the emission occurs before the spin polarization decays, leading to larger $\bar{\rho}_c$ in sub-100-nm GaSe nanoslabs. The decrease of $\tau_0$ is unlikely due to an enhancement of radiative recombination rates with dimensionality and confinement \cite{andreani1991} given that the bulk Bohr exciton radius of GaSe is only $\sim$5 nm. The linear dependence of $\tau_0 \propto d_L$ can result from the propagation of exciton-polaritons \cite{aaviksoo1991} or non-radiative surface recombination \cite{aspnes1983} (see Supplementary Information). The PL measurements presented here only probe excitons and carriers near the $\Gamma$ point. Nonetheless, the distinct PL dynamics at room and cryogenic temperatures suggest that there is significant non-radiative recombination due to inter-valley scattering between the $\Gamma$ and M valleys \cite{capozzi1993}. 

The reduced mixing of valence bands makes GaSe and other group-III monochalcogenides such as GaS and InSe \cite{kuroda1980} promising materials for devices relying on the ability to generate and maintain high degrees of spin polarization. In these materials, the hole spin relaxation rate is expected to decrease significantly compared to that in III-V semiconductors because the EY relaxation mechanism is suppressed by the reduced mixing of distant valence bands. This hypothesis is supported by the quantitative exciton and spin relaxation rates obtained by fitting the experimental polarized PL dynamics as described in Supplementary Information. Nanoscale multi-layer GaSe or InSe are, in principle, less challenging to synthesize and integrate with micro- and nano-fabrication processes compared to atomic membranes of TMDs. Additionally, non-centrosymmetric group-III monochalcogenides allow for more direct optical control of light-matter coupling beyond the monolayer regime.

\subsection*{Methods}
\textbf{Sample growth and characterization.} %The crystalline 
GaSe single crystals  are grown from stoichiometric amounts of high purity (7N) precursors using the vertical Bridgman method \cite{mandal2008a}. X-ray powder diffraction results confirmed the hexagonal $\epsilon$-GaSe structure with $a$ = 3.743 $\AA$ and $c$ = 15.916 $\AA$. Inductively coupled plasma emission mass spectroscopy (ICPMS) was carried out to determine the stoichiometry of these samples. These GaSe samples are stoichiometric with 50.001\% Ga (atomic) and 49.997\% Se (atomic).

\textbf{Sample preparation.} Like graphene/graphite and layered transition metal dichalcogenides, GaSe can be mechanically exfoliated into atomically thin few-layer crystals. Atomically thin and nanometer thick GaSe crystals are mechanically exfoliated from a Bridgman-grown crystal and deposited onto a silicon substrate with a 90 nm \ce{SiO2} layer. Sample thickness is measured by atomic force microscopy (Supplementary Fig. S1). Samples are mounted in vacuum on a copper cold finger attached to an optical liquid helium flow cryostat for all experiments. 

\textbf{Optical pumping.}
GaSe nanoslabs are optically excited by a 2-ps pulsed laser with a varying central wavelength between 560 to 610 nm from a synchronously pumped optical parametric oscillator. The laser beam is focused through a microscope objective (numerical aperture N.A. = 0.28) to an area of about 80 $\mu$m$^2$ on the sample. The wave vector of the pump is along the crystal $c$-axis (the surface normal), i.e. the electric field vector $\vec{E}$ is orthogonal to the $c$-axis ($\vec{E} \perp c$). The maximum deviation from normal incidence is 8$^{\circ}$ in air and $\sim 2.5 ^{\circ}$ in the crystal for this objective. The polarization and pump flux ($P$) of the pump laser is controlled by liquid-crystal-based devices without mechanical moving parts. We estimate the photoexcitation density to be from $\approx 2\times10^{16}$ cm$^{-3}$ to $3.4\times10^{17}$ cm$^{-3}$ ($2.7\times10^{-10}$ cm$^{-2}$ per layer) considering the absorption coefficient at 2.1 eV ($\approx10^3$ cm$^{-1}$ for $\vec{E} \perp c$) and Fresnel loss from reflection. The photoexcited carrier density is below the Mott transition \cite{pavesi1989,capozzi1993} of direct excitons occurring near e-h pair densities of $4\times10^{17}$ cm$^{-3}$.

\textbf{Time-integrated and time-resolved photoluminescence spectroscopy.}
We characterize decays of optically excited carrier by measuring reflected photoluminescence (PL) propagating along the $c$-axis through a standard microscopy set-up. PL spectra are observed using an imaging spectrometer (linear dispersion 1 nm/mm, focal length 750 mm, grating 300 grooves/mm) equipped with a liquid-nitrogen cooled CCD. To characterize the temporal evolution of the exciton/carrier population, we monitored time-dependent photoluminescence near the band edge using a streak camera system.

\textbf{Polarized time-dependent photoluminescence.}
To better understand the spin relaxation of photoexcited excitons (electrons/holes) in GaSe crystalline nanoslabs, we analyzed time-dependent polarization properties of PL under nonresonant circularly and linearly polarized 2-ps pulsed pump. The spin-lattice relaxation time $\tau_s$ of the photoexcited carriers can be determined from co-circularly and cross-circularly polarized time-dependent PL. 

\subsection*{Acknowledgments}
This work is supported by Michigan State University and by NSF through DMR-09055944. This research has made use of the W. M. Keck Microfabrication Facility. We thank Dat Do, Brage Golding, Bhanu Mahanti, and Carlo Piermarocchi for comments and discussions.

\subsection*{Author contributions}
Y.T. led the optical measurements and data analysis, assisted by W. X.. K.M. grew the materials. C.W.L. conceived the project and developed the theoretical model. J.A.M. and C.W.L. supervised the project and wrote the paper with input from all authors.

%\subsection*{Additional information}
%The authors declare no competing financial interests.

\end{document}

\title{Supplementary Information for \\ ``Near Unity Optical Spin Polarization in GaSe Nanoslabs"}

\author{Yanhao Tang}
\author{Wei Xie}
\affiliation{Department of Physics and Astronomy, Michigan State University, East Lansing, MI 48824, USA}
\author{Krishna C. Mandal}
\affiliation{Department of Electrical Engineering, University of South Carolina, Columbus, SC 29208}
\author{John A. McGuire}
\author{C. W. Lai}
\email{cwlai@laigrp.com}
\affiliation{Department of Physics and Astronomy, Michigan State University, East Lansing, MI 48824, USA}

%\date{\today}

%\begin{abstract}
%\end{abstract}

\maketitle

\section{Supplementary Experimental Figures}
\subsection{Optical Microscope and AFM Images}
\begin{figure}[H]
\centering
\includegraphics[width=0.8 \textwidth]{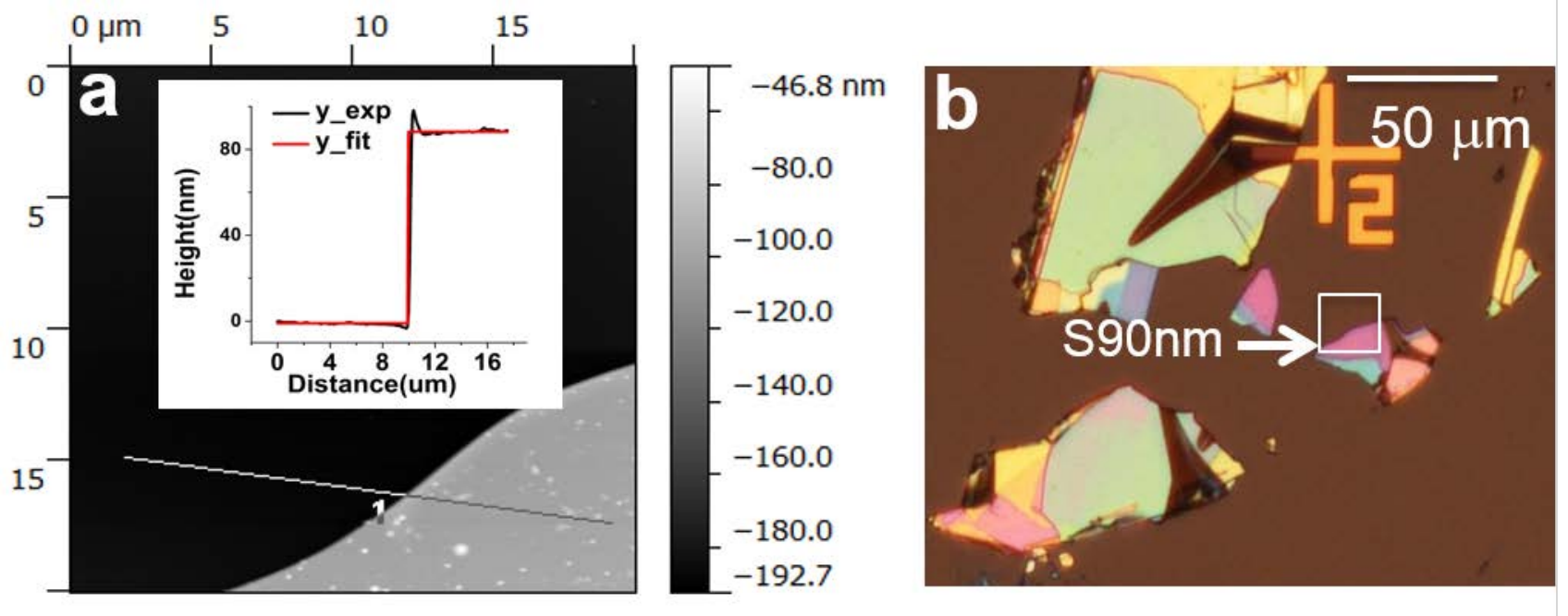}
\caption{\textbf{Optical identification of GaSe nanoslabs.}\textbf{a}, An AFM image and cross-sectional profile (inset) of a 90-nm thick GaSe nanoslab (S90nm). AFM images are obtained using an atomic force microscope, Asylum Research Cypher S. \textbf{b}, An optical microscope image of GaSe samples on a 90-nm SiO2/Si substrate. The boxed area around the sample S90nm is the AFM scanning area shown in \textbf{a}.}
\label{fig:optical_images}
\end{figure}

\subsection{Quantum Efficiency}
\begin{figure}[H]
\centering
\includegraphics[width=0.55 \textwidth]{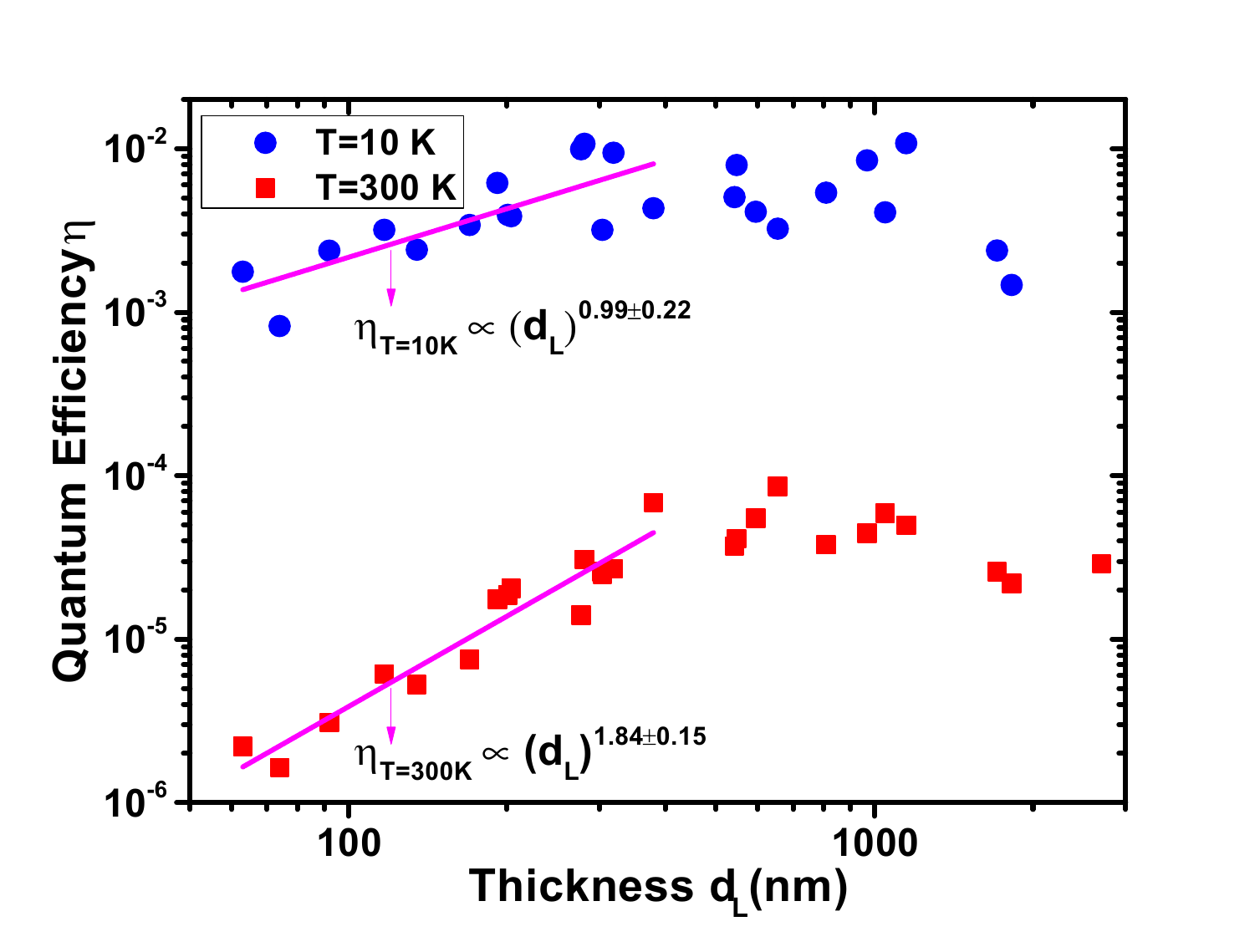}
\caption{\textbf{PL quantum efficiency.} Quantum efficiency of luminescence as a function of thickness of GaSe nanoslabs at T = 300 and 10 K. Here, quantum efficiency is defined as the ratio between the luminescence emission flux and optical absorption flux per layer. Optical absorption flux is determined using experimentally measured reflectance for each sample and tabulated absorption coefficient $\alpha= 1.1\times10^3$ cm$^{-1}$ measured by Le Toullec et al. and Adachi et al. \cite{le-toullec1977,*le-toullec1980,*piccioli1977,*adachi1992}. Optical collection and detection efficiency is measured by passing a 633-nm laser beam with known power through the optical set-up and spectrometer. The emission flux is then calculated by including Fresnel reflection loss at the surface assuming angularly isotropic emission.}
\end{figure}

\subsection{Polarized Time-integrate PL at T = 10 K}
\begin{figure}[H]
\centering
\includegraphics[width=0.55 \textwidth]{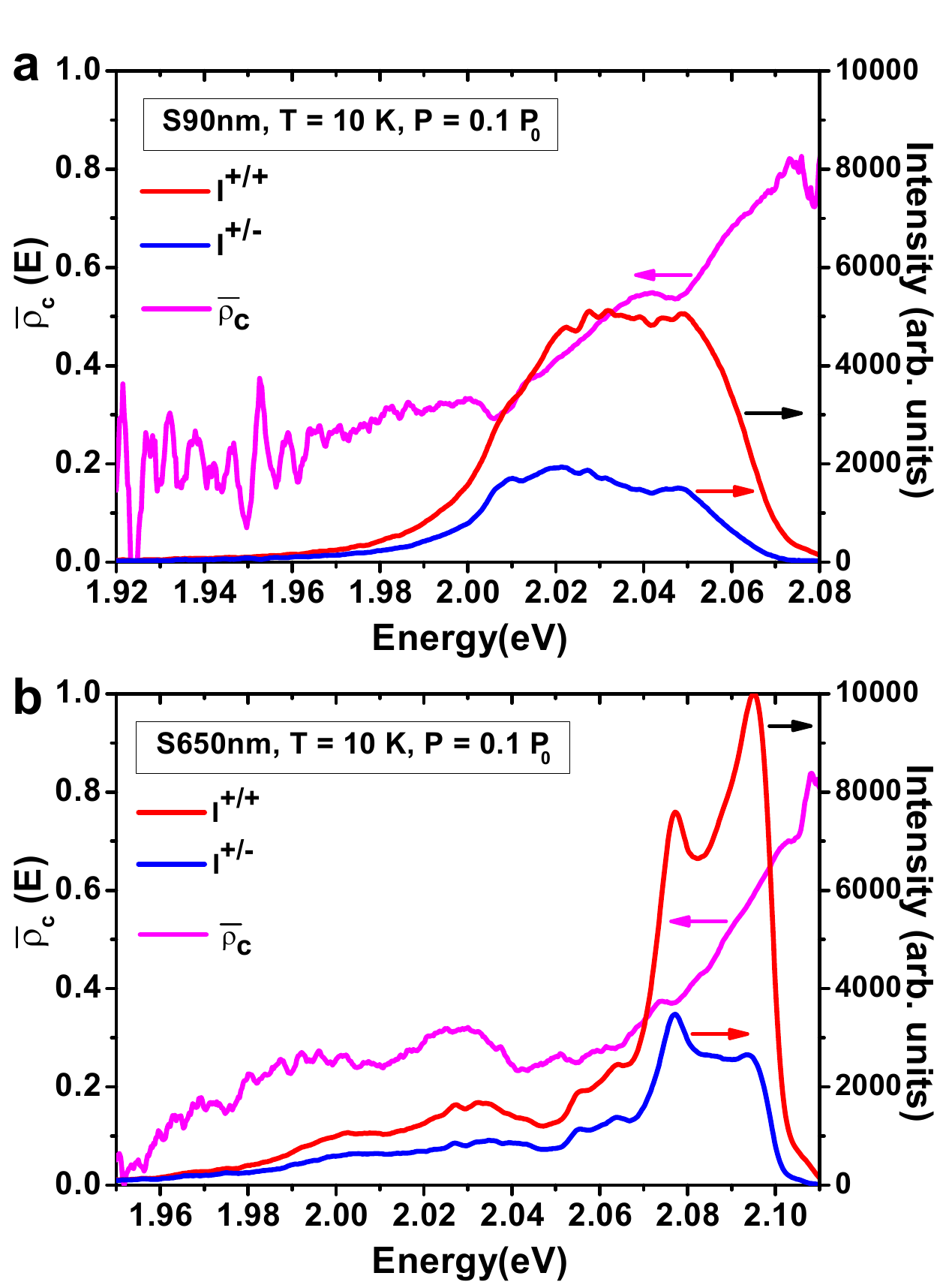}
\caption{\label{fig:PL10KP010}\textbf{Polarized time-integrated spectra at $P$ = $P_0$ at T = 10 K.} Time-integrated PL spectra [$I^{+/+}(E)$(co-circular, red) and $I^{+/-}(E)$(cross-circular, blue)] and degree of circular polarization $\bar{\rho}_c(E)$ of \textbf{a}, S90nm and \textbf{b}, S650nm under $\sigma^+$ excitation at pump flux $P$ = 0.1 $P_0$.
}
\end{figure}

\subsection{Excitation Energy Dependence}
\begin{figure}[H]
\centering
\includegraphics[width=0.55 \textwidth]{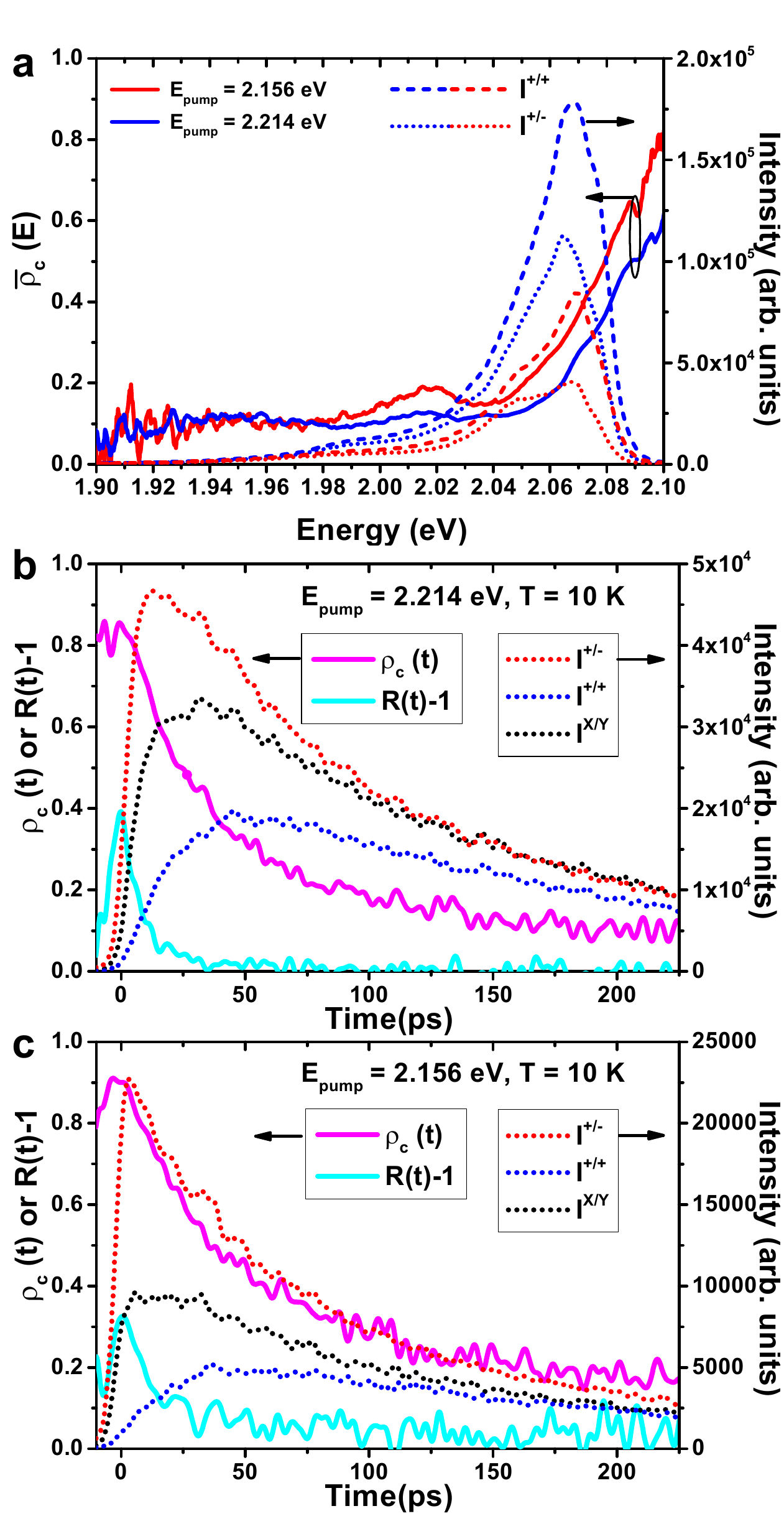}
\caption{\textbf{Polarized PL spectra and dynamics under optical excitations at 2.156 eV and 2.214 eV.} \textbf{a}, Time-integrated PL spectra [$I^{+/+}(E)$(co-circular, dashed lines) and $I^{+/-}(E)$(cross-circular, dotted lines)] and degree of circular polarization $\bar{\rho}_c(E)$ (solid lines) of 540-nm thick (S540) GaSe samples under $\sigma^+$ excitation at pump flux $P$ = 0.5 $P_0$, where $P_0$ = $2.6\times10^{14}$  cm$^{-2}$ per pulse. Blue (Red) lines are for $E_{pump}$ = 2.214 eV (2.156 eV). \textbf{b}, Time-dependent PL intensity [$I^{+/+}(t)$, $I^{+/-}(t)$, and $I^{X/Y}(t)$ (dotted red, blue, and black, respectively)], degree of circular polarization $\rho_c(t)=\frac{I^+(t)-I^-(t)}{I^+(t)+I^-(t)}$ (solid magenta), and $R(t)-1= \frac{I^{+/+}(t)+I^{+/-}(t)}{I^{X/X}(t)+I^{X/Y}(t)}-1$ (solid cyan) under excitation $E_{pump}$ = 2.214 eV at $P$ = 0.5 $P_0$. \textbf{c}, Same as \textbf{b}, but for excitation $E_{pump}$ = 2.156 eV.
}\label{fig:ex560-570nm}
\end{figure}

\begin{figure}[H]
\centering
\includegraphics[width=1.0 \textwidth]{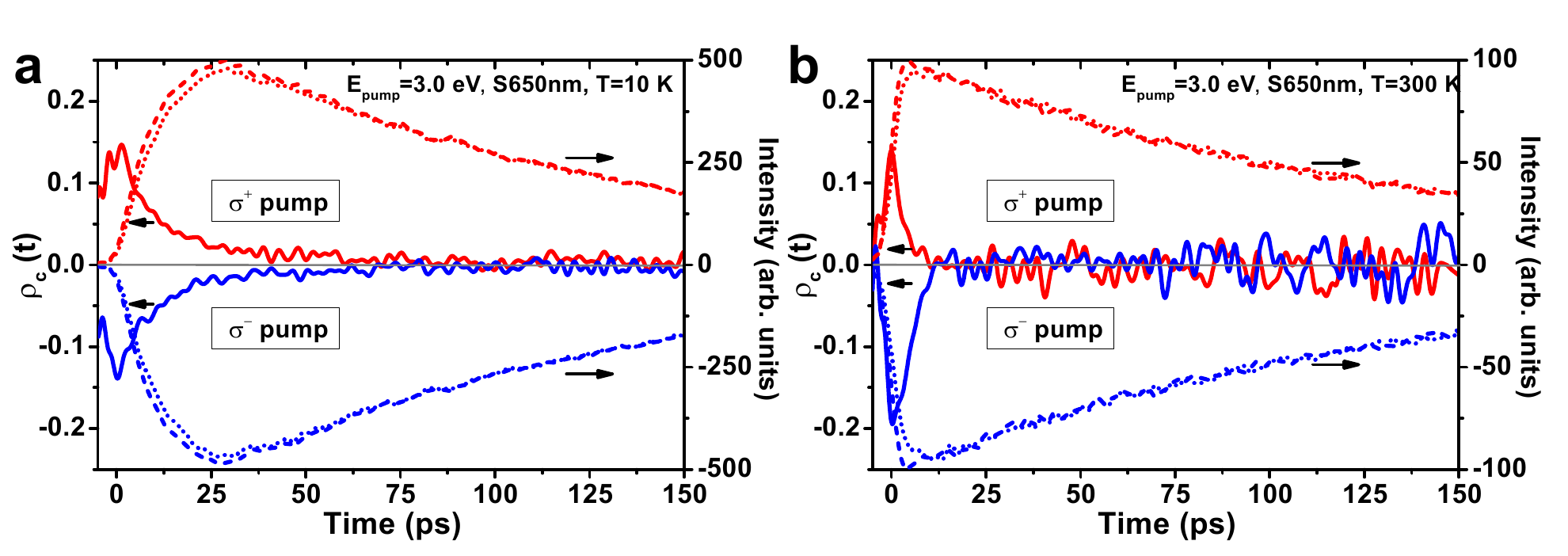}
\caption{\textbf{Polarized time-dependent PL under optical excitation at 3 eV.} Time-dependent PL intensity $I^{\pm/\pm}(t)$ (co-circular) and $I^{\pm/\mp}(t)$ (cross-circular) (dashed and dotted lines, respectively), degree of circular polarization $\rho_c(t)$ (solid lines) under excitation $E_{pump}$ = 3 eV. Red and blue lines are for $\sigma^+$ and $\sigma^-$ excitation, respectively. \textbf{a}, T = 300 K, and \textbf{b}, T = 10 K.  The decay of $\rho_c(t)$ is slower at T = 10 K than that at T = 300 K ($\approx$7 ps vs. 3 ps) as is the PL rise time ($\approx$18 ps vs. 5 ps). 
}\label{fig:ex413nm}
\end{figure}

\newpage
\section{Band Structure and Spin-Orbit Interaction}\label{sec:bandstructure}
The III-VI semiconducting compounds GaS, GaSe, and GaTe (MX) all form layered crystals. The bonding between the layers is weak, resulting in the easy cleavage of these crystals. In the case of GaS and GaSe, each layer consists of four planes of atoms in the sequence X-M-M-X and belong to the space group $D^1_{3h}$. Different stacking orders of hexagonal layers result in formation of four commonly known polytypes: $\beta$ ($D^4_{6h}$), $\epsilon (D^1_{3h})$, rhombohedral $\gamma (C^5_{3v})$, and $\delta$. Symmetry-dependent optical transitions and spin dynamics near the band edge can be sensitive to polytypes \cite{brebner1967}. Here, we study $\epsilon-$GaSe which has an ABA (Bernal) stacking order and belongs to space group $D_{3h}^1-P\bar{6}m2$, which lacks an inversion center between adjacent layers. 

Gallium selenide (GaSe) is a layered semiconductor the nonlinear optical properties of which have been extensively studied \cite{abdullaev1972,*fernelius1994}. Bulk GaSe is generally regarded as an indirect band gap semiconductor with nearly degenerate conduction band minima at the $\Gamma$ and $M$ points of the Brillouin zone. The indirect transition to the $M$ point lies $\sim 10-20$ meV below the direct gap ($\Gamma$), and thus is nearly resonant with the direct exciton transitions at the $\Gamma$-point (exciton binding energy 20-30 meV) \cite{brebner1967,*aulich1969,*mercier1973,*mooser1973,*schluter1976,*le-toullec1977,*le-toullec1980,*sasaki1981,*capozzi1993}. The conduction-band and valence-band structures of $\epsilon$-GaSe are illustrated in Fig. 1 in the main text. The uppermost valence band (UVB) near the $\Gamma$ point has Se $p_z$ symmetry, while the lowest conduction band mostly has Ga $s$ symmetry. Valence bands are split by crystal-field anisotropy and spin-orbit interaction, leading to two bands with Se $p_x, p_y$ symmetry about 1.2 and 1.6 eV below the UVB maximum. In the thin crystals used in the experiments presented here, the absorption slightly above the band gap is attributed primarily to the direct transition with negligible absorption from the indirect transition. 

The essential optical selection rules near $k=0$ ($\Gamma$-point) were explained by Mooser using an exciton (two-particle) picture \cite{mooser1973,*minami1976,*gamarts1977,*ivchenko1977} and are detailed below. The selection rules for direct-gap excitons in $\epsilon$-GaSe including spin-orbit coupling are better illustrated in the two-particle (exciton) picture as shown in Fig. 1c in the main text. The upper level $\Gamma_4$ corresponds to a total exciton spin $S$ = 0, and experiences a splitting $\Delta_1 \approx 2$ meV due to the electron-hole exchange interaction \cite{mooser1973}. The states $\Gamma_3$ and $\Gamma_6$ correspond to a total exciton spin $S$ = 1, and $S_z$= 0, $\pm1$ and are nearly degenerate (energy splitting $\Delta \, \approx \, 0$). These states are thus labeled by the indices 0 and $\pm1$. The $\Gamma_4$ state can be excited by light with $\vec{E} \, \parallel \, c$. For optical excitation with wave vector $\vec{k} \parallel c$, $\Gamma_6$ ($S_z \, = \pm1$) states can be excited by circularly polarized light with $\vec{E} \perp c$, whereas the $\Gamma_3$ state is optically inactive \cite{mooser1973, *gamarts1977,*ivchenko1977}. Earlier studies of GaSe suggested a high degree of optical orientation in GaSe could be achieved under nearly resonant excitation at low temperatures. Through steady-state measurements, 
Gamarts et al. \cite{gamarts1977,*ivchenko1977} demonstrated optical orientation and alignment of excitons in GaSe by showing luminescence with circular polarization above 90\% under steady-state circularly polarized optical excitation \emph{in resonance} with direct excitons at \emph{cryogenic} temperatures.

Next, we discuss the relationship between the $\vec{k}$ = 0 ($\Gamma$ point) conduction-band and valence-band states of GaSe in comparison with energy bands in zinc-blende and wurtzite structures. An essential difference between the potential that an electron experiences in an ideal hexagonal GaSe or wurtzite lattice and that in a zinc-blende lattice (e.g. GaAs) is due to the crystal field from sites beyond the next nearest neighbors. In a wurtzite structure (e.g. CdSe and ZnO), the crystal field leads to a crystal-field splitting ($\Delta_{cr}$) at the $\Gamma$ point between a doublet ($\Gamma_5$) and a singlet ($\Gamma_1$), corresponding to the triplet states of zenc-blende ($\Gamma_{15}$). In GaSe, the uppermost $p_z$-like valence band is separated from the lower $p_{x,y}$-like valence bands by about 1.4 eV \cite{sasaki1975a,*sasaki1975b,*sasaki1981}. The actual energy splitting among these three uppermost valence bands in GaSe is due to the combined effect of spin-orbit and crystal-field perturbations. Using a quasi-cubic model \cite{hopfield1960,*kuroda1980}, we can write the wavefunction for each band as a linear combination of $p_x$, $p_y$, and $p_z$ and spin functions, provided that the coupling to other conduction and valence bands is neglected.
 
The energy difference of the split-off valence bands can be expressed in terms of the spin-orbit interaction ($\Delta_{so}$) and crystal-field splitting ($\Delta_{so}$) as:
\begin{align}
E_{AB} &= \frac{\Delta_{so}+\Delta_{cr}}{2} \, - \sqrt{(\frac{\Delta_{cr}+\Delta{so}}{2})^2-\frac{2}{3}\Delta_{cr} \Delta{so}} \, ,\nonumber \\
E_{AC} &= \frac{\Delta_{so}+\Delta_{cr}}{2} \, + \sqrt{(\frac{\Delta_{cr}+\Delta{so}}{2})^2-\frac{2}{3}\Delta_{cr} \Delta{so}} \, .
\end{align}
For $\epsilon$-GaSe, the uppermost valence band is assigned to $B$ and the second and third bands are assigned to $A$ and $C$, respectively. We obtain $E_{BA} = E_B-E_A = 1.27$ eV and $E_{AC}=E_A-E_C=0.32$ eV using $\Delta_{so}$= 0.44 eV and $\Delta_{cr}$= -1.39 eV measured by Sasaki et al. \cite{sasaki1975a,*sasaki1975b}. 

The $\epsilon$-GaSe crystal has a $D_{3h}$ point-group symmetry, the symmetry of a single layer of GaSe. At the center of the Brillouin zone, the conduction band has $\Gamma_{4}$ ($s$-like) symmetry and the $B$, $A$, and $C$ valence bands have $\Gamma_{1}$, $\Gamma_{5}$, and $\Gamma_{6}$ ($p$-like) symmetry, respectively. The polarization vectors $\vec{E} \perp c$ and $\vec{E} \parallel c$ belong to the $\Gamma_{4}$ and $\Gamma_{6}$ representations, respectively. Considering the transition between uppermost valence band ($B$) and conduction band without spin, the direct product $\Gamma_1 \times \Gamma_4$ belongs to the representation $\Gamma_4$ of $D_{3h}$; therefore, only the direct transition $\Gamma_1 \rightarrow \Gamma_4$ for $\vec{E} \parallel c$ is allowed according to the orbital symmetries. Taking into account spin by going into the double group $\bar{D}_{3h}$, the direct transitions then occur between valence and conduction bands with the following symmetries:
\begin{align}
\Gamma_4 &\rightarrow \Gamma_8 \, (s\text{-like conduction band}) \nonumber \\
\Gamma_1 &\rightarrow \Gamma_7 \, (B, p_z\text{-like valence band}) \nonumber \\
\Gamma_5 &\rightarrow \Gamma_7 + \Gamma_9 \, (A, p_{x,y}\text{-like valence band}) \nonumber \\
\Gamma_6 &\rightarrow \Gamma_8 + \Gamma_9 \, (C, p_{x,y}\text{-like valence band})  
\end{align}
The direct product $\Gamma_7 \times \Gamma_8 = \Gamma_3 + \Gamma_4 + \Gamma_6$ then contains the representations for both $\vec{E} \perp c$ and $\vec{E} \parallel c$. This means that for $\vec{E} \perp c$  the optical transition between the $B$ valence and conduction bands is weakly allowed only if spin-orbit coupling is taken into account, while for $\vec{E} \parallel c$ all optical transitions are dipole-allowed irrespective of spin-orbit coupling.

\subsection{Direct Excitons}
The direct transition from the valence and conduction band spinors at the $\epsilon$-GaSe band edge belongs to $\Gamma_7$ and $\Gamma_8$ of the double group $\bar{D}_{3h}$. Four $s$-like direct-gap exciton states are possible in GaSe \cite{mooser1973,*schluter1976,*gamarts1977,*ivchenko1977,*sasaki1981}:
\begin{equation}
\Gamma^{(s)}_{X}=\Gamma_7 \times \Gamma_8 \times \Gamma_1 = \Gamma_4 + \Gamma_3 + \Gamma_6.
\end{equation}
The corresponding wavefunctions of the three valence bands are as follows \cite{mooser1973}:
\begin{align}
\Gamma_4 &= \left| S \right> + \alpha_4 \left| T \right>, \nonumber \\
\Gamma_3 &= \left| T \right>, \nonumber \\
\Gamma_6 &= \left| T \right> + \alpha_6 \left| S \right>.
\end{align}
where $\left| S \right>$ and $\left| T \right>$ represent singlet and triplet states, respectively. The coefficients $\alpha_4 \approx \alpha_6 \approx (\Delta_{so}/E_{BA}) \approx 0.35$. The $\Gamma_1$ component mixes about 10\% of the $\Gamma_5$ state, consistent with the ratio of the oscillator strength (absorbance) between $\vec{E} \perp c$ and $\vec{E} \parallel c$. The upper level $\Gamma_4$ corresponds to a total exciton spin $S$ = 0, and experiences a splitting $\Delta_1 \approx 2$ meV due to electron-hole exchange interaction \cite{mooser1973}. The states $\Gamma_3$ and $\Gamma_6$ correspond to a total exciton spin $S$ = 1, and $S_z$= 0, $\pm1$ and are nearly degenerate (energy splitting $\Delta \, \approx \, 0$). These states are thus labeled by the indices 0 and $\pm1$. The $\Gamma_4$ state can be excited by light with $\vec{E} \, \parallel \, c$. For optical excitation with wave vector $\vec{k} \parallel c$, $\Gamma_6$ ($S_z \, = \pm1$) states can be excited by circularly polarized light with $\vec{E} \perp c$, whereas the $\Gamma_3$ state is optically inactive \cite{gamarts1977,*ivchenko1977}.

\section{Spin relaxation mechanism}
\subsection{Overview}
Spin relaxation mechanisms in semiconductors depend on characteristics of the band structure, such as the energy versus momentum dispersion and spin-splitting due to spin-orbit interaction \cite{luttinger1955,*kane1957}. In III-V semiconductors such as GaAs, the spin orientation of holes is lost in a period comparable to the momentum relaxation time ($\tau_p$) owing to the strong coupling between the angular momentum and quasimomentum of holes. The relaxation of \emph{electron} spin in semiconductors is usually analyzed in terms of three mechanisms \cite{dyakonov1984,*pikus1984,*wu2010,*boross2013}: Dyakonov-Perel (DP) \cite{dyakonov1972}, Elliott-Yafet (EY) \cite{elliott1954,*yafet1963}, and Bir-Aronov-Pikus (BAP) \cite{bir1973,*pikus1974,*bir1975,*aronov1983}. In general, the spin-orbit coupling can yield intra- or interband mixing of orbital bands \cite{boross2013}. %

Although hole spin is typically lost rapidly, the same fundamental mechanisms determine hole spin relaxation with EY typically being the dominant mechanism for hole spin relaxation in III-V semiconductors due to the near degeneracy of light- and heavy-hole bands and the split-off band \cite{yu2005,*shen2010}.

\paragraph{\textbf{DP mechanism}} In noncentrosymmetric III-V compounds, the lack of inversion symmetry leads to a spin splitting of the conduction band for $k\neq0$, which is described by the presence of a $k^3$ term in the conduction-electron spin Hamiltonian \cite{zerrouati1988}. This splitting is equivalent to the presence in the crystal of an effective magnetic field inducing the precession of electron spins. This yields the DP mechanism wherein momentum scattering need not change the spin direction but changes the direction of the effective magnetic field resulting in loss of spin polarization as each carrier precesses about a different effective field. Rapid momentum scattering produces motional narrowing and so a slowing of the decay of the spin polarization. %
In the case of thermalized electrons, the spin relaxation rate depends on the kinetic energy $\epsilon \sim k_BT$ and is given by 
\begin{equation}
\frac{1}{T_1(\epsilon)} = Q \alpha^2 \frac{(k_BT)^3}{\hbar^2 E_g} \, \tau_p(n) \, ,
\end{equation}
where $\tau_p$ is the momentum scattering/relaxation time and can be dependent on the carrier density $n$. The parameter $Q$ depends on the scattering process of electrons and $\alpha$ is characteristic of the band structure reflecting the $k^3$ term of the conduction electron spin Hamiltonian. The temperature dependence of the spin relaxation rate is %
also subject to the variation of $\tau_p$ with temperature.

\paragraph{\textbf{EY mechanism}}
In crystals with inversion symmetry, intraband coupling is prohibited, and the mixing of other bands (typically valence) with the conduction band results in the electron states no longer being pure spin states but remaining degenerate. This gives rise to the EY mechanism whereby momentum scattering changes the spin state, so that rapid momentum scattering yields rapid loss of spin polarization \cite{elliott1954,*yafet1963}. Spin relaxation can arise from the elastic scatting of the electrons/holes by acoustic and optical phonons as well as by impurities. The spin relaxation rates due to the EY mechanism have been calculated for different momentum relaxation processes \cite{pikus1984,*wu2010}. In general, the spin-lattice relaxation time $T_1$ is also proportional to the momentum relaxation time $\tau_p$:
\begin{equation}
\frac{1}{T_1(\epsilon)} \propto \frac{1}{\tau_p(n,\epsilon)}.
\end{equation}

\paragraph{\textbf{BAP mechanism}}
In \emph{p}-type semiconductors, spin relaxation may result from electron scattering by holes with simultaneous exchange interaction \cite{bir1973,*pikus1974,*bir1975,*aronov1983,*zerrouati1988}. BAP \cite{bir1973,*pikus1974,*bir1975,*aronov1983} proposed a mechanism considering the efficiency of such an electron-hole exchange process. They found that two terms, involving, respectively, free and bound holes, appear in the formula describing the spin relaxation rate in a nondegenerate semiconductor:
\begin{equation}
\frac{1}{T_1}= \frac{2}{\tau_0} N_A a_B^3 \frac{v_e}{v_B} \left| \frac{N_P}{N_A} |\psi(0)|^4 + \frac{5}{3} \left(1-\frac{N_P}{N_A} \right) \right|,
\end{equation}
where $v_B=\hbar/\epsilon_e a_B$ is the exciton Bohr velocity and $v_e=(3 K_B T/m_e^*)^{1/2}$ for thermalized electrons. $\tau_0$ is calculated from the exchange splitting $\Delta_{X}$ of the exciton ground state, $a_B$ is the exciton Bohr radius, $N_P/N_A$ represents the acceptor degree of ionization, $|\psi(0)|^2$ is the Sommerfeld factor, and $N_P$ is the density of free holes. 

\subsection{Spin relaxation in GaSe}
The situation is different in GaSe, which has an uppermost $p_z$-like valence band weakly coupled to other valence bands. Most importantly, the layer structure leads to a large crystal field that produces an orbitally non-degenerate upper valence band separated at the $\Gamma$ point by more than 1 eV from the nearest bands (valence bands derived primarily from Se $p_x,\,p_y$ orbitals). While the aforementioned mechanisms are generally discussed in the context of \emph{electron} spin relaxation, this is simply a consequence of the extremely fast relaxation of holes in III-V semiconductors. In contrast, the reduced mixing of valence band states in GaSe compared to III-V semiconductors should result in correspondingly reduced EY contributions to the hole spin relaxation. In the absence of more detailed band-structure calculations, we cannot determine quantitatively the spin relaxation rates due to the EY or DP mechanisms near the $\Gamma$ point. However, whichever dominates is expected to result in much slower hole spin relaxation than in, e.g., GaAs \cite{fishman1977,*zerrouati1988,*dyakonov2008,*amand2008,*wu2010}. The relative contribution of each spin relaxation mechanism may change with temperature, doping, and photoexcited carrier density. For photoexcited carrier density above $10^{16}$ cm$^{-3}$ as studied in undoped GaSe here, we neglect spin relaxation due to the BAP mechanism. The momentum scattering time $\tau_p(n)$ is expected to decrease with increasing carrier density $n$. Therefore, we expect $T_1$ to increase (decrease) with $n$ if spin relaxation is dominated by EY (DP). In our experiments, at T= 10 K, we found the initial decay of $\rho_c$ is characterized by $\tau'_s\propto n^{-0.23}$. This suggest the EY mechanism plays a larger role than the DP mechanism in the initial spin relaxation in GaSe at low temperature.

\subsection{Spin relaxation of non-thermalized electrons and holes}
When the electron (hole) lifetime is much longer than the energy relaxation time, the majority of electrons (holes) in the conduction (valence) band is thermalized under stationary excitation. Spin-momentum correlation is generally lost during thermalization. However, during the thermalization process and before the average spin polarization is completely lost, polarized \emph{hot} photoluminescence may be observed. This can yield a higher degree of circular polarization at high energies in the steady-state spectrum. In GaSe, the initial carrier cooling to the band edge occurs in the sub-ps to sub-10-ps range \cite{nusse1997}. This fast energy and momentum relaxation (cooling) is demonstrated by the sub-10-ps PL rise time that is nearly independent of PL emission energy. Additionally, the PL decay is similar across the spectrum throughout the detection range (data not shown). Consequently, the measured time-dependent and spectrally integrated PL is dominated by the emission near the band edge exciton even in the initial 10-20 ps. 

\section{A Rate-Equation Model: Exciton Dynamics and Spin Relaxation}

\iffalse
\todo[inline]{
1. Develop model and solve rate equations for exciton spin relaxation \cite{damen1991,*bar-ad1992,*sham1993,*maialle1993,*vinattieri1994,*amand1994,*munoz1995,*ivchenko1995}. \\
2. Check: Is there any spin-splitting \cite{vina1996,*le-jeune1998,*ciuti1998,*de-leon2001} observed in time-resolved and spectrally-resolved streak images? \\
3. Discuss spin relaxation mechanisms at T = 10 K: DP and BAP mechanisms, exciton spin relaxation \\
4. Cite discussions on DP, BAP, and other spin relaxation mechanisms in III-VI layered materials? \\
- Simplify discussions of DP, EY, and BAP mechanisms\\
- Estimate spin relaxation rate due to DP mechanism \\
}
\fi

\begin{figure}[H]
\centering
\includegraphics[width=0.85 \textwidth]{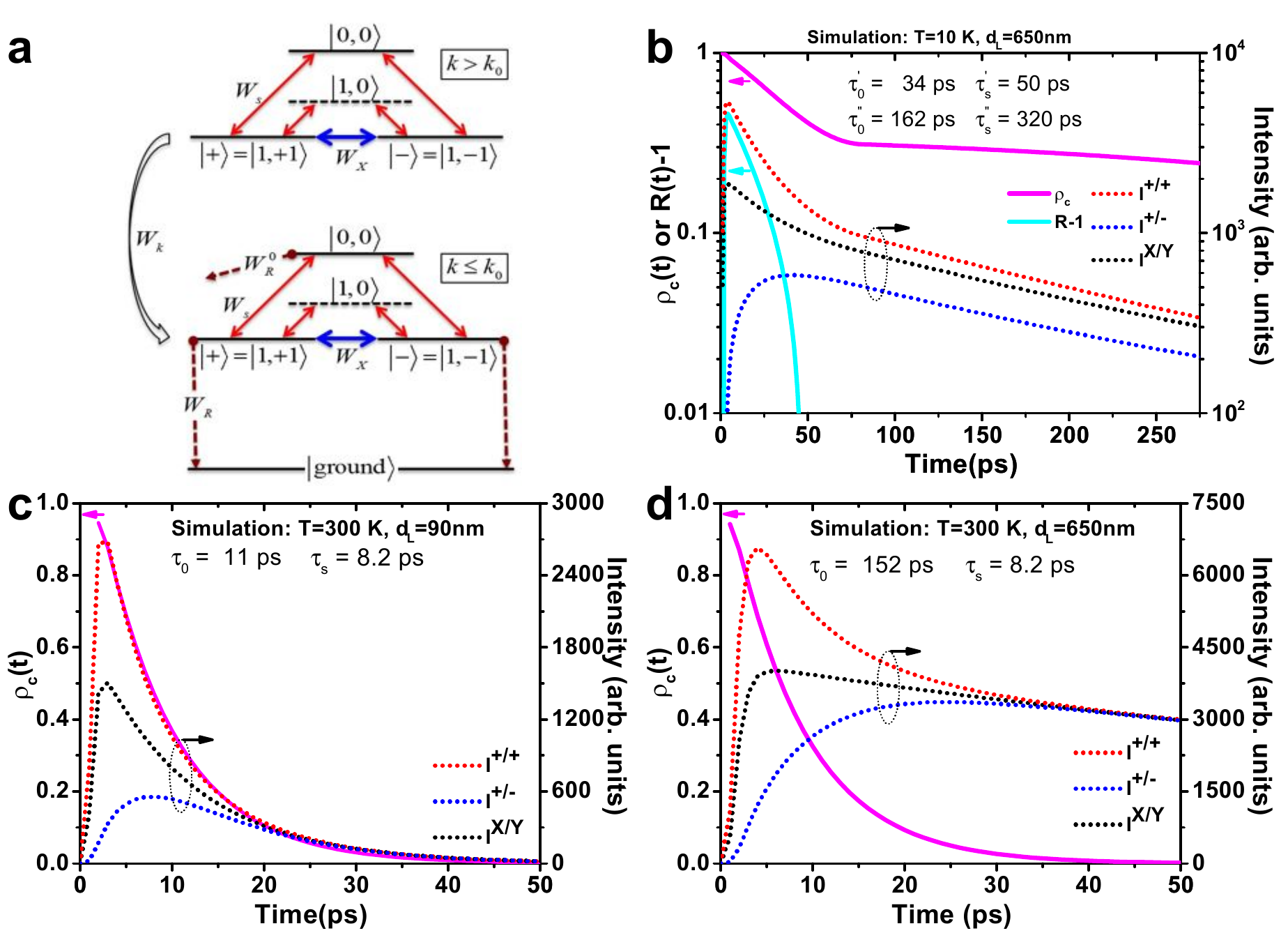}
\caption{\textbf{a}, Schematic of the model for the exciton dynamics. \textbf{b-d}, Calculated polarized PL dynamics for \textbf{b}: T = 10 K, $d_L$ = 650 nm, \textbf{c}: T = 300 K, $d_L$ = 90 nm, and \textbf{d}: T = 300 K, $d_L$ = 650 nm.The parameters used in the calculation are listed in Table \ref{tb:model}.
}\label{fig:spinmodel} 
\end{figure}

The dynamics of resonantly excited non-thermal excitons in quasi-two-dimensional systems such GaAs-based quantum-well structures has been shown to be affected by several different physical processes \cite{damen1991,*sham1993,*maialle1993,*vinattieri1994,*wang1995,*munoz1995,*baylac1995,*ivchenko1995,*amand1997,*le-jeune1998,*vina1999,*amand2008}: (1) the momentum relaxation of excitons, (2)  the spin relaxation of excitons, and (3) the enhanced radiative recombination and propagation of exciton polaritons. For non-resonantly photo-excited carriers, it might be necessary to consider the contribution to dynamics from free carriers, in particular at high temperature. Here, we calculate the population of excitons in various spin and momentum states in GaSe as described in Sec. \ref{sec:bandstructure} using a simplified model (Fig. S\ref{fig:spinmodel}a). The model is adapted from a unified model for resonantly excited excitons in GaAs-based quantum wells \cite{vinattieri1994} and should be valid only for resonant photoexcitation at low temperature where the contributions from free carriers is negligible. Nevertheless, such a simple model can reproduce most of the experimental photoluminescence polarization properties and dynamics at both cryogenic temperature (T = 10 K) and room temperature.

We first consider the case when the excitons are photoexcited in the non-radiative $\left|+1k\right>$ state ($K_\parallel>K_0$), where $K_\parallel$ is the in-plane momentum of excitons and $K_0$ is the phonon momentum. Exciton spin-flip (with rate $W_X$) transfers population between $\left|+\right>=\left|1,1\right>$ and $\left|-\right>=\left|1,-1\right>$ states, while electron/hole spin-flip (rate $W_s$ indistinguishable for electron and hole) populates the dipole-inactive $\left|10\right>=\left|1,0\right>$ non-radiative (dark) state, and the singlet $\left|00\right>=\left|0,0\right>$ state (dipole-active for $E\parallel c$). Following Vinattieri et al. \cite{vinattieri1994}, we divide the manifold of $K_\parallel$ states into two sets, one for nearly zero $K_\parallel$ and the other for finite large $K_\parallel$ states. Each set includes four exciton states ($\left|\pm\right>$, $\left|1\right>$, and $\left|0\right>$). Absorption and emission of acoustic phonons induce transitions between these two sets. We consider only spin-conserving transitions with an effective scattering rate $W_k$. To simplify the model, we neglect any thermal factors associated with the spin-flip rates of excitons and electron/hole.   

The time-dependent population in each state is given by a set of coupled equations:
\begin{equation}
\frac{d}{dt}N_i=M_{ij} \, N_j + G(t) \, \delta_{+1k,i} \,\, ,
\end{equation}
where $N_i$ is the column vector $(N_{+1}, N_{-1}, N_{10}, N_{00}, N_{+1k}, N_{-1k}, N_{10k}, N_{00k})$ and $M$ is a $8\times8$ matrix. $N_{+1}, N_{-1}, N_{10}, N_{00}$ are populations of $K_\parallel \lesssim K_0$ states $\left|+\right>$, $\left|-\right>$, $\left|10\right>$, and $\left|00\right>$, respectively. $N_{+1k}, N_{-1k}, N_{10k}, N_{00k}$ are corresponding $K_\parallel>K_0$ states. $M$ is the following matrix:
\begin{equation}
M=
\begin{bmatrix}
A & C \\
D & B
\end{bmatrix},
\end{equation}
where $A$, $B$, $C$, and $D$ are the following $4\times4$ matrices:
\begin{equation}
A=\left[
\begin{smallmatrix}
-(W_r+W_X+W_s+W_{kp}) & W_X & W_s/4 & W_s/4 \\
W_X & -(W_r+W_X+W_s+W_{kp}) & W_s/4 & W_s/4 \\
W_s/2 & W_s/2 & -(W_s/2+W_{kp}) & 0 \\
W_s/2 & W_s/2 & 0 & -(W_{r}^{0}+W_s/2+W_{kp})
\end{smallmatrix}\right], \nonumber\\
\end{equation}
\begin{equation}
B=\left[
\begin{smallmatrix}
-(W_X+W_s+W_{km}) & W_X & W_s/4 & W_s/4 \\
W_X & -(W_X+W_s+W_{km}) & W_s/4 & W_s/4 \\
W_s/2 & W_s/2 & -(W_s/2+W_{km}) & 0 \\
W_s/2 & W_s/2 & 0 & -(W_s/2+W_{km})
\end{smallmatrix}\right], \nonumber\\
\end{equation}

\begin{equation}
C=
\begin{bmatrix}
W_{km} & 0 & 0 & 0 \\
0 & W_{km} & 0 & 0 \\
0 & 0 & W_{km} & 0 \\
0 & 0 & 0 & W_{km} \\
\end{bmatrix}, \,
D=
\begin{bmatrix}
W_{kp} & 0 & 0 & 0 \\
0 & W_{kp} & 0 & 0 \\
0 & 0 & W_{kp} & 0 \\
0 & 0 & 0 & W_{kp} \\
\end{bmatrix}. \nonumber
\end{equation} 
$W_r$ and $W_r^0$ are the radiative recombination rates for $\left|\pm 1/2\right>$ and $\left|0\right>$, respectively. The radiative recombination rate $W_r^0$ for state $\left|0\right>$ is set to 30 $W_r$ based on the relative absorption coefficients between $E \parallel c$ and $E \perp c$ light \cite{le-toullec1977}. $W_X$ and $W_s$ are the exciton and electron/hole spin-relaxation rates. The acoustic-phonon scattering rates $W_{kp}$ and $W_{km}$ are defined as by Vinattieri et al. \cite{vinattieri1994} as 
\begin{align}
W_{kp}=W_k \, \textrm{exp}\left[-\frac{\Gamma_h}{k_B\,T}\right], \nonumber\\
W_{km}=W_k \, \left(1-\textrm{exp}\left[-\frac{\Gamma_h}{k_B\,T}\right]\right),
\end{align}
where $W_k$ is the effective scattering rate with phonons, and $\Gamma_h$ is associated with the homogenous linewidth. The spin-flip rates of electrons and holes can not be distinguished in this model because spin-flip of electron and holes results in identical transitions within the model.

Next we consider the time-dependent PL under linearly polarized excitation. In this case, the photoexcited electrons and holes are assumed to be equally distributed over their respective spin states, i.e. $\left|\pm 1/2\right>$, during the initial 2-ps laser excitation. The non-geminate (bimolecular) formation of excitons then produces an initial exciton population distributed equally over the three triplet exciton states $\left|\pm\right>$ and $\left|1\right>$. We consider two specific excitation conditions: (1) circularly polarized excitation ($\sigma^+$), and (2) linearly polarized excitation ($\sigma^X$). The total time-dependent populations of $\left|+\right>$ and $\left|-\right>$ for these two cases are $I^+(t)$ and $I^X(t)$, respectively. Then the ratio $R(t)=I^+(t)/I^X(t)$ will decrease from approximately 1.5 to 1 because the populations in three triplet states eventually become nearly equal through spin relaxation under $\sigma^+$ excitation.

There are five parameters in the model ($W_r$, $W_X$, $W_s$, $W_k$, $\Gamma_h$); however, we can obtain quantitative information about these rates by fitting the experimental polarized PL dynamics with this model (see Fig. S\ref{fig:spinmodel} and Table \ref{tb:model}). The value of $\Gamma_h$ is not available but in principle can be measured independently. For simplicity, we set $\Gamma_h\,=\,2\,k_B\,T$, independent of temperature. $W_{nr}$ is determined by the thickness-dependent quantum efficiency and is limited to less than 10 ps$^{-1}$, i.e. the non-radiative recombination lifetime $\tau_{nr}$ is limited to more than 0.1 ps (100 fs). $W_r$ is then largely determined by the decay of population (PL decay). 

The polarized PL dynamics under circularly polarized excitation is affected by both $W_X$ and $W_s$. However, the sum of time-dependent PL under circularly polarized $\sigma^\pm$ excitation and PL under linearly polarized $\sigma^{X/Y}$ excitation are independent of $W_X$ because $W_X$ does not change the total population in the two radiative states. By fitting experimental results, we find that polarized PL dynamics is dominated by $W_s$ at cryogenic temperatures (T = 10 K), and by $W_X$ at room temperature. Note that exciton polarization relaxation in GaAs-based quantum-well structures have also been found to be dominated by exchange spin flip at room temperature \cite{tackeuchi1990,*tackeuchi1999}, while the polarized PL dynamics at cryogenic temperatures is either dominated by electron/hole spin flip or a combination of exciton exchange and single electron/hole spin flip \cite{damen1991,*bar-ad1992}.  

At T = 10 K, the time-dependent degree of circular polarization [$\rho_c(t)$] exhibits bi-exponential decay. The corresponding time-dependent PL under linearly polarized excitation also decays bi-exponentially. Additionally, the ratio $R(t)$ decays to 1 within about 20 ps. These experimental results can only be reproduced by considering the effects of non-resonantly excited non-thermal carriers. In principle, the experimental PL decay is a characteristic of a thermalized exciton/carrier distribution near the band edge and provides no information about the dynamics of nonthermal excitons. However, polarized PL dynamics can still reveal the effect of spin relaxation of nonthermal excitons as these excitons cool and scatter into the region $K_\parallel\leq K_0$. We expect the spin-flip rate of electrons and holes to be highly dependent on their energy and momentum due to a combination of the Dyakonov-Perel and Elliot-Yafet spin relaxation mechanisms. Consequently, the non-resonantly photoexcitation carriers can experience a high spin-flip rate during the thermalization process \cite{vina1999}. To simulate such an energy- and momentum-dependent spin relaxation rate without considering detailed non-equilibrium carrier distributions, we assign a spin-flip rate that decreases linearly from about $\Omega \times W_s$ to 1 $W_s$ during the thermalization period $\tau_{th}$. The thermalization time can be determined by fitting to the polarized PL dynamics at low temperature ($\tau_{th}\sim$ 80 ps at T = 10 K). After including such a phenomenological parameter $\Omega=500$ and $\tau_{th}=80$ ps in the model, we reproduce quantitatively $\rho_c(t)$, $I^{+/+}(t)$, $I^{+/-}(t)$, and $I^{X/Y}(t)$, and qualitatively [$R(t)-1$] (Fig. S\ref{fig:spinmodel}b). The calculated PL dynamics agree with the experimental results shown in Fig. 5 in the main text. $\Omega$ and $\tau_{th}$ are expect to increase with photoexcitation energy. We can thus increase $\Omega$ and $\tau_{th}$ to fit polarized PL dynamics with higher excitation energy as those shown in Fig. S\ref{fig:ex560-570nm} and Fig. S\ref{fig:ex413nm}. 

At room temperature (T = 300 K), the sharp decrease of the co-circularly polarized PL component is accompanied by a similar rise of the cross-circularly polarized component. The sum of the two components and the time-dependent PL under linearly polarized excitation show a mono-exponential decay. In this case, the spin (polarization) dynamics is determined by $W_X$. Another unique feature of room-temperature PL is the linear dependence of the PL decay time ($\tau_0$) on thickness ($d_L$) for nanoslabs thinner than about 800 nm. Here, we attribute the decrease of PL decay time to a thickness-dependent effective radiative recombination rate $W_r(d_L)$ (``exciton-polariton scheme"). Such a linear dependence on sample thickness can result from the propagation of exciton-polariton along the $c-$axis \cite{aaviksoo1991}, i.e. $W_r={v'_g}/{d_L}+W_r^{int.}$, where $v'_g=4.2\times10^5$ cm/s is the effective group velocity and $W_r^{int}$ the intrinsic radiative recombination rate in the bulk as determined experimentally. The calculated polarized PL dynamics and time-averaged (stationary) $\bar{\rho}_c$ in samples of thickness 90 nm and 650 nm using the model with a thickness-dependent $W_r(d_L)$ (Fig. S\ref{fig:spinmodel}c-f) are in agreement with the experimental results shown in Fig. 3 in the main text. Despite the good agreement using an exciton-polariton picture, we caution that the fundamental radiative recombination rate of the excitons \cite{andreani1991,*andreani1995,*citrin1992,*citrin1993,*citrin1995} as well as the role of non-radiative recombination in GaSe nanosalbs are still open problems. The cooling and ionization of nonresonantly excited \emph{hot} carriers and the subsequent exciton formation, trapping, and exciton-exciton and exciton-carrier scattering \cite{taylor1987,*minami1990,*yao1983,*capozzi1993,*nusse1997} are also not considered in the model. The exciton-carrier scattering process \cite{mercier1975,*mercier1975a,*capozzi1983,*capozzi1983a,*capozzi1993} can also play a role in the luminescence dynamics. Moreover, the spontaneous emission rate in nanoscale and atomically thin layered semiconductors can differ from that in bulk and even be controlled as shown for monolayer \ce{MoS2} deposited on a photonic crystal nanocavity \cite{gan2013}. Finally, we note that radiative and non-radiative recombination have been investigated in thin crystal films \cite{liu1975,*sumi1981,*aaviksoo1991,*knoester1992,*bjork1994b,*bjork1995} and III-V quantum wells \cite{feldmann1987,*hanamura1988,*damen1990,*andreani1991,*deveaud1991,*deveaud1991a,*srinivas1992,*citrin1992,*colocci1993,*deveaud1993,*sermage1993,*sermage1993a,*citrin1993,*bjork1994a,*citrin1995,*citrin1995a,*andreani1995}. In quantum wells, radiative recombination has also been studied as a function of carrier density \cite{srinivas1992,*eccleston1992} and temperature\cite{knox1985,*gurioli1991,*gurioli1994,*martinez-pastor1993,*piermarocchi1996}. 

There are several experimental results this simple model cannot produce quantitatively. For example, the polarized PL dynamics is nearly independent of sample thickness at T = 10 K, in contrast to the PL dynamics at room temperature. In addition, the quantum efficiency decreases quadratically with thickness ($d_L$ at T = 300 K, but almost linearly with $d_L$ at T = 10 K for samples thinner than about 400 nm. To model these effects, it is necessary to include other carrier scattering and non-radiative recombination processes such as the surface recombination discussed below. 

\subsection{Surface recombination}\label{sec:surface}
Here, we consider surface recombination \cite{nelson1978,*aspnes1983,*ahrenkiel1991,*ahrenkiel1992} to model the decrease of PL decay time with thickness at T = 300 K. The non-radiative recombination rate induced by surface recombination is expected to decrease significantly at cryogenic temperature as demonstrated by the enhanced quantum efficiency measured at T = 10 K. By fitting the PL dynamics at T = 10 K with negligible $W_{nr}$, we determined $W_r=0.006$ as before. Assuming the radiative recombination rate decreases linearly with temperature as for excitons in a quasi-two-dimensional systems, we deduced $W_r=2.0\times10^{-4}$ ps$^{-1}$ for calculations at T = 300 K. In this case, the room-temperature PL decay is dominated by non-radiative recombination, particularly in thin samples. To fit experimental polarized PL dynamics at T = 300 K, we use a non-radiative recombination rate $W_{nr}(d_L)\,=\, [S/d_L+W_{nr}(\infty)]\times A\exp(-E_a/k_B\,T)$, where $S$ is the surface recombination velocity and $W_{nr}(\infty)\approx0.005$ is a phenomenological parameter introduced to emulate other non-radiative recombination processes in bulk and to fit PL dynamics in thick samples ($d_L$ $>$ 800 nm). $S=4.2\times10^5$ cm/s is associated with the velocity measured experimentally. Such a surface recombination velocity would be comparable to that in pristine GaAs which is know to have a high surface recombination velocity compared to other semiconductors such as GaN, InP, and Si. $A=1.215$ is a normalization constant, and $E_a=5$ meV is the thermal activation energy determined by the relative quantum efficiency at T = 10 K and 300 K. Here, for simplicity, we use the same $E_a$ for all non-radiative recombination and scattering processes that results in \emph{loss} of optically active carriers.

The coupled equations for the surface recombination scheme have the following $A$ and $B$ matrices with additional $W_{nr}$ terms:
\begin{equation}
A=\left[
\begin{smallmatrix}
-(W_r+W_X+W_s+W_{kp}+W_{nr}) & W_X & W_s/4 & W_s/4 \\
W_X & -(W_r+W_X+W_s+W_{kp}+W_{nr}) & W_s/4 & W_s/4 \\
W_s/2 & W_s/2 & -(W_s/2+W_{kp}+W_{nr}) & 0 \\
W_s/2 & W_s/2 & 0 & -(W_{r}^{0}+W_s/2+W_{kp}+W_{nr})
\end{smallmatrix}\right], \nonumber\\
\end{equation}

\begin{equation}
B=\left[
\begin{smallmatrix}
-(W_X+W_s+W_{km}+W_{nr}) & W_X & W_s/4 & W_s/4 \\
W_X & -(W_X+W_s+W_{km}+W_{nr}) & W_s/4 & W_s/4 \\
W_s/2 & W_s/2 & -(W_s/2+W_{km}+W_{nr}) & 0 \\
W_s/2 & W_s/2 & 0 & -(W_s/2+W_{km}+W_{nr})
\end{smallmatrix}\right]. \nonumber\\
\end{equation}

The parameters used for calculations are listed in Table \ref{tb:model_surface}. The calculated time-dependent polarized PL curves are essentially the same as those using the exciton-polariton scheme. However, the surface recombination scheme produces qualitatively the decreasing quantum efficiency with decreasing $d_L$ and increasing temperature. 
 
%\newpage

\begin{table}[H]
\caption{Modeling Parameters: Exciton-Polariton Scheme}
\centering
\begin{tabular}{|l|p{1.8cm}|p{1.8cm}|p{1.8cm}|p{1cm}|p{1.5cm}|p{1.5cm}|p{1.5cm}|}
\hline
Parameter     &$W_r$(ps$^{-1}$) &$W_X$(ps$^{-1}$)    &$W_s$(ps$^{-1}$)   &$\Omega$ &$\tau_{th}$(ps) &$\Gamma_h$  &$W_k$(ps$^{-1}$)\\ \hline
10 K, 650 nm  & 0.006           &$<10^{-5}$ & 0.001             & 500     & 80             & $2k_B\,T$ & 1.0            \\ \hline
300 K, 650 nm & 0.00123         & 0.0625             &$<10^{-5}$& 500     & 80             & $2k_B\,T$ & 1.0            \\ \hline
300 K, 90 nm  & 0.0089          & 0.0625             &$<10^{-5}$& 500     & 80             & $2k_B\,T$ & 1.0            \\ \hline
\end{tabular}\label{tb:model}
\end{table}

\begin{table}[H]
\caption{Modeling Parameters: Surface Recombination Scheme}
\centering
\begin{tabular}{|l|p{1.8cm}|p{1.8cm}|p{1.8cm}|p{1.8cm}|p{1cm}|p{1.5cm}|p{1.5cm}|p{1.5cm}|}
\hline
Parameter     &$W_r$(ps$^{-1}$) &$W_{nr}$(ps$^{-1}$)&$W_X$(ps$^{-1}$)    &$W_s$(ps$^{-1}$)   &$\Omega$ &$\tau_{th}$(ps) &$\Gamma_h$  &$W_k$(ps$^{-1}$)\\ \hline
10 K, 650 nm  & 0.006           & $<10^{-4}$        &$<10^{-5}$ & 0.001             & 500     & 80             & $2k_B\,T$ & 1.0            \\ \hline
300 K, 650 nm & 0.0002          & 0.0117            & 0.0625             &$<1.0\times10^{-5}$& 500     & 80             & $2k_B\,T$ & 1.0            \\ \hline
300 K, 90 nm  & 0.0002          & 0.0472            & 0.0625             &$<1.0\times10^{-5}$& 500     & 80             & $2k_B\,T$ & 1.0            \\ \hline
\end{tabular}\label{tb:model_surface}
\end{table}

\newpage
\section*{Supplementary References}